\PassOptionsToPackage{table,dvipsnames}{xcolor} % 放在导言区最前面
\documentclass[sigconf,natbib=true]{acmart}
\AtBeginDocument{%
  \providecommand\BibTeX{{%
    \normalfont B\kern-0.5em{\scshape i\kern-0.25em b}\kern-0.8em\TeX}}}

\usepackage{algorithm}
\usepackage{algorithmic}
\usepackage{multirow}
\usepackage{balance}
\usepackage{subfigure}
\usepackage{graphicx}
\usepackage{mdframed}
\usepackage{tcolorbox}

\usepackage{enumitem}

\usepackage{amsmath}
\usepackage{amsthm}

\newtheorem{definition}{Definition}

\usepackage{ragged2e}
\usepackage{colortbl}
\usepackage[table]{xcolor}

\definecolor{lightblue}{RGB}{173,216,230}
\definecolor{lightgreen}{RGB}{144,238,144}

\newcommand{\ie}{\emph{i.e., }}
\newcommand{\eg}{\emph{e.g., }}

\newcommand{\nosection}[1]{\vspace{2pt}\noindent\textbf{#1.}}

\newcommand{\modelname}{\textbf{TPAD}}

\newcommand{\RAMP}{\textbf{TPAD}}

\begin{document}

\title{Distilling Transitional Pattern to Large Language Models for Multimodal Session-based Recommendation}
% \mys How about this title: Recurrently Distill Multi-order Transition to Large Language Models for Session-based Recommendation
% reasons: 1) I feel "distill on xxx" is weired, actually, "distill sth to sth" makes more sense; 2) transition can just be singular rather than plural;  

\begin{abstract}

Session-based recommendation (SBR) predicts the next item based on anonymous sessions.
Traditional SBR explores user intents based on ID collaborations or auxiliary content.
To further alleviate data sparsity and cold-start issues, recent Multimodal SBR (MSBR) methods utilize simplistic pre-trained models for modality learning but have limitations in semantic richness.
Considering semantic reasoning abilities of Large Language Models (LLM), we focus on the LLM-enhanced MSBR scenario in this paper, which leverages LLM cognition for comprehensive multimodal representation generation, to enhance downstream MSBR.
Tackling this problem faces two challenges: i) how to obtain LLM cognition on both transitional pat-
terns and inherent multimodal knowledge, ii) how to align both features into one unified LLM, minimize discrepancy while maximizing representation utility.
To this end, we propose a multimodal LLM-enhanced framework TPAD, which extends a distillation paradigm to decouple and align transitional patterns for promoting MSBR.
TPAD establishes parallel Knowledge-MLLM and Transfer-MLLM, where the former interprets item knowledge-reflected features and the latter extracts transition-aware features underneath sessions.
A transitional pattern alignment module harnessing mutual information estimation theory unites two MLLMs, alleviating distribution discrepancy and distilling transitional patterns into modal representations.
Extensive experiments on real-world datasets demonstrate the effectiveness of our framework.
%

% \footnote{{Our {implementation code} will be released later at https://github.com/XXXX/HMTA}.}. 

\end{abstract}

\author{Jiajie Su}
\affiliation{Zhejiang University
\institution{}
\country{Hangzhou,China}
}
\email{sujiajie@zju.edu.cn}

\author{Qiyong Zhong}
\affiliation{Zhejiang University
\institution{}
\country{Hangzhou,China}
}
\email{youngzhong@zju.edu.cn}

\author{Yunshan Ma}
\affiliation{Singapore Management University	
\institution{}
\country{Singapore}
}
\email{ysma@smu.edu.sg}

\author{Weiming Liu}
\affiliation{Zhejiang University
\institution{}
\country{Hangzhou,China}
}
\email{21831010@zju.edu.cn}

\author{Chaochao Chen}
\affiliation{Zhejiang University
\institution{}
\country{Hangzhou,China}
}
\email{zjuccc@zju.edu.cn}

\author{Xiaolin Zheng}
\affiliation{Zhejiang University
\institution{}
\country{Hangzhou,China}
}
\email{xlzheng@zju.edu.cn}

\author{Jianwei Yin}
\affiliation{Zhejiang University
\institution{}
\country{Hangzhou,China}
}
\email{zjuyjw@cs.zju.edu.cn}

\author{Tat-Seng Chua}
\affiliation{National University of Singapore	
\institution{}
\country{Singapore}
}
\email{dcscts@nus.edu.sg}

\renewcommand{\shortauthors}{Jiajie Su et al.}


% \ccsdesc[300]{Collaborative filtering}


\maketitle

\setlength{\floatsep}{4pt plus 4pt minus 1pt}
\setlength{\textfloatsep}{4pt plus 2pt minus 2pt}
\setlength{\intextsep}{4pt plus 2pt minus 2pt}
\setlength{\dbltextfloatsep}{3pt plus 2pt minus 1pt}
\setlength{\dblfloatsep}{3pt plus 2pt minus 1pt}
\setlength{\abovecaptionskip}{3pt}
\setlength{\belowcaptionskip}{2pt}
\setlength{\abovedisplayskip}{2pt plus 1pt minus 1pt}
\setlength{\belowdisplayskip}{2pt plus 1pt minus 1pt}

\section{Introduction}

% 介绍session-based rec
Session-based recommendation (SBR) \cite{wu2019session,xia2021co,hou22core} is a practical and challenging scenario, where users' profiles and long-term behaviors are unavailable due to privacy policies or non-logged-in nature.
Thereby, exploring short-term sequential patterns within anonymous sessions remains the crux of SBR.
Initial attempts rely on ID-based methods~\cite{rendle2010factorizing,tan2016improved,wang2020global,li2023ultrare}, which are restricted by sparsity and cold-start problems \cite{cold-start-sess,chen2023knowledge}.
To mitigate these problems, content-based methods~\cite{lai2022attribute,jin2023dual,liu2024enhancing} utilize auxiliary information, \eg item attributes, knowledge graphs, to enhance representations with rich semantics.
Recently, several studies investigate \textbf{Multimodal SBR} (MSBR) pipelines, \ie MMSBR \cite{zhang2023beyond} and DIMO \cite{zhang2024disentangling}, which fuse various modalities to portray comprehensive user interests.
However, the performance of existing MSBR models is bottlenecked by the low quality of multimodal representations which are often extracted by inferior pre-trained models.

% characterize dynamic user intents by disentangling synergistic interplay across modalities.
%
% However, existing MSBR leverages simplistic pre-trained models to generate modal embeddings, where the narrow representation capability leads to inadequate semantic depiction.
%
\begin{figure}[t]
\centering
\includegraphics[width=1.0\linewidth]{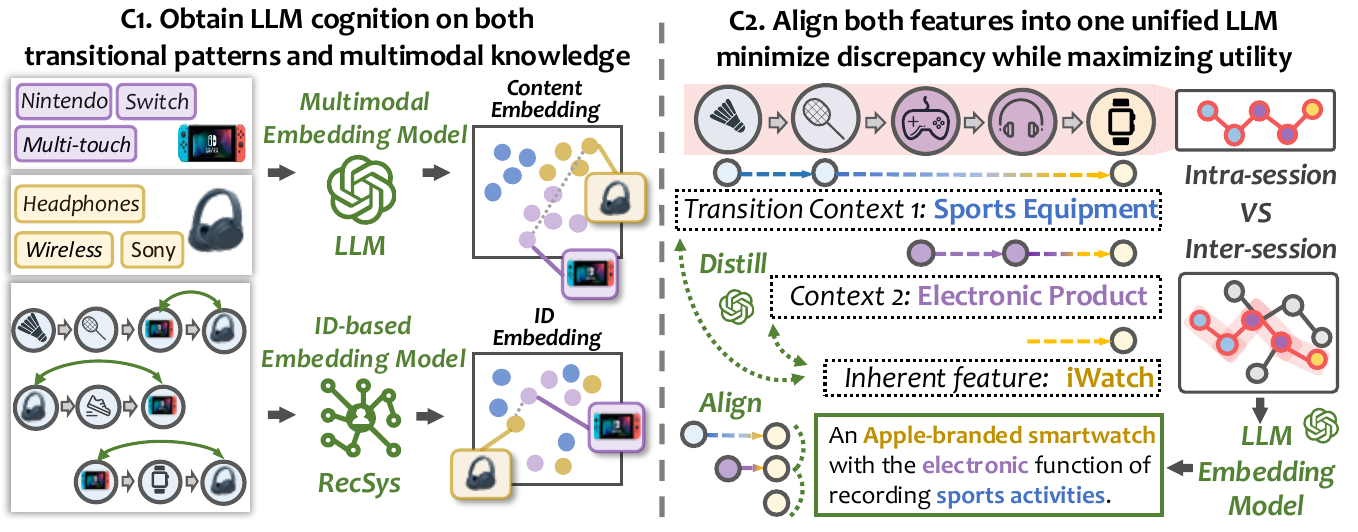}
% \vspace{-0.25 in}
\caption{Motivation of TPAD. C1 reveals inconsistency of two spaces and C2 uncovers complex transitions.} 
\label{story}
\vspace{-0.1 in}
\end{figure}

% Given extensive knowledge and sophisticated reasoning capabilities of Large Language Models (LLMs), unlocking the potential of LLMs in the MSBR scenario becomes inevitable. 
The recent breakthrough of Large Language Models \cite{luo2024mmevol,zhou2024few,zhou2025dreamdpo,feng2024fine} sheds light on solving such bottlenecks given their extensive knowledge and sophisticated reasoning capabilities, where two paradigms obtain growing attention: \textit{LLM-centric} and \textit{LLM-enhanced} approaches. 
% 现有的LLM+REC有两类做法，各自的问题是什么
%
Specifically, \textit{LLM-centric} approaches~\cite{zhang2023recommendation,sun2024large,guo2024integrating,wang2024re2llm} directly reform LLMs as recommenders with designed instructions.
Considering the expansive decision space involved in generating the next item and the accompanying issue of hallucinations, these works commonly reformulate the recommendation task as a binary classification problem (\eg TALLRec \cite{bao2023tallrec}), or a multiple-choice question answering problem (\eg LLaRA \cite{liao2024llara}).
Although LLMs can work under well-curated datasets and toy settings, they inevitably suffer from accuracy drop and inference inefficiency in practical scenarios, when the input token length increases due to session growth ~\cite{hu2024enhancing} or the all-ranking\footnote{Take the whole item set as candidates and give a ranking list.} setting~\cite{ren2024representation}.
Given these limitations, researchers start to prioritize another branch of methods, \ie \textit{LLM-enhanced} approaches \cite{hou2023learning,liu2024llm,liu2024practice}, which treat LLMs as powerful embedding models to interpret item content, producing expressive representations to downstream recommenders.
These studies leverage LLM's superior capability in representation learning rather than preference modeling, circumventing problems caused by lengthy sequences and delegating the all-ranking tasks to conventional recommenders.
For instance, NoteLLM2 \cite{zhang2024notellm-2} integrates language models and vision encoders to construct multimodal embeddings for item-to-item recommendation, while Molar \cite{luo2024molar} generates unified item representations with a post-alignment mechanism.
Even though these works pioneer the study of LLM-enhanced paradigm, they still encounter significant constraints when adapted to MSBR due to the following challenges:

\nosection{C1: How to obtain LLM cognition on both transitional patterns and inherent multimodal knowledge} 
% what is transitional patterns; what is multimodal knowledge
Multimodal knowledge represents item intrinsic features, while transitional patterns indicate sequential behavioral cues suggesting ID collaborations.
% we need to maximize their collaborative effects 
% but they are inconsistency
As shown in Figure \ref{story}.C1, obvious distributional inconsistency exists between the \textit{knowledge-reflected feature} space and \textit{transition-aware feature} space. 
Thus, discriminating their heterogeneity is essential to facilitate holistic representation learning in LLM.
Previous studies \cite{wei2024towards,zhang2024notellm,luo2024molar} employ LLM to only model the knowledge-reflected features, or directly generate both features in one shared space.
Such coarse-grained modeling (i) leads to an incomplete LLM perception of \textit{dynamic} and \textit{static} item characteristics, (ii) homogeneous representation generation and poor generalizability.
% potentially result in negative transfer~\cite{ren2024representation} that harms pre-existing semantic LLM cognition.
% potentially resulting in negative transfer or semantic bias.
%
Multimodal knowledge and transition patterns are intertwined in sessions, making it difficult to disentangle them in LLM comprehension.

% which motivates us to interpret item representations as a unified form of \textit{knowledge-reflected} and \textit{transition-aware} features.

% Previous studies \cite{wei2024towards,zhang2024notellm,luo2024molar} employ multimodal LLM to depict knowledge-reflected features based on item self-content, but fall short of disentangling transition-aware features derived from multifaceted transitional patterns.

% Recognizing the complexity of intertwined relations in sessions, decoupling transitional patterns with inherent multimodal features and aligning them in the semantic space remain difficult.

% \nosection{C2: How to endow LLM with reasoning ability to distill transitional patterns into representation generation}
\nosection{C2: How to align both features into one unified LLM, minimize discrepancy while maximizing representation utility}
% 1. complex transitional patterns   2. absorb transition-aware features while preserving knowledge-reflected features, keep a balance
% 
\textit{On the one hand}, session data contains diverse transitional patterns, which exhibit hierarchical cooperation of item contents.
As shown in Figure~\ref{story}.C2, users' evolving intents within a session are externalized as transitional patterns in distinct semantic contexts.
And item transitions exhibit both intra- and inter-session \cite{xia2021co,wang2023multi}, further complicating the extraction of patterns.
\textit{On the other hand}, due to distributional bias between knowledge-reflected and transition-aware features, roughly incorporate both potentially results in negative transfer~\cite{ren2024representation} that harms pre-existing semantic LLM cognition.
% problem in previous work: 1、not discriminative 2、only local/intra-order
Previous works \cite{ma2024triple,ye2024harnessing} include entire sessions into the prompt, which (i) ignore discriminating diverse patterns existing in intra-session local and inter-session global transitions, (ii) fail to selectively absorb transition-aware features while preserving knowledge-reflected features.
Realizing a balanced integration in LLM of transitional patterns into item multimodal representation remains challenging.
%
% Enhancing LLM's ability in comprehensive representation generation remains challenging.

% The multifaceted semantics expressed in multi-order transition collectively form items' \textit{transition-aware features}.
%
% Thus, overlooking the layered structure of transitions hinders the accurate alignment between transition-aware and knowledge-reflected features.

% conclusion
% Thus, empowering LLM with layered comprehension of multi-order transitions and integrate them into unified multimodal representations is essential.

To tackle these challenges, we follow the LLM-enhanced paradigm and propose a novel framework that realizes \underline{T}ransitional \underline{Pa}ttern \underline{D}istillation (TPAD) to Large Language Model for MSBR.
\modelname~proposes a multimodal distilling paradigm to disentangle and align transitional patterns into multimodal representation learning, ultimately enhancing the performance of downstream MSBR.
%
% C1: we propose K-MLLM, T-MLLM, and TPA alignment
\textbf{To address C1}, we propose a dual-tower MLLM structure, \ie \textit{Knowledge-MLLM (K-MLLM)} and \textit{Transfer-MLLM (T-MLLM)}, which disentangles the learning process of static and dynamic item features. 
\textit{K-MLLM} is guided by item modality reasoning prompts to cognize \textit{knowledge-reflected features} within static modalities, while \textit{T-MLLM} is instructed by session transition reasoning prompts to extract \textit{transition-aware features} within dynamic session context.
%% and \textit{Transitional Pattern Alignment (TPA)}
%
% C2: we propose the multi-stage distillation paradigm
\textbf{To overcome C2}, we develop the \textit{Transitional Pattern Alignment (TPA)} and the \textit{distillation paradigm}, which achieve comprehensive incorporation of transitional patterns into multimodal representations.
The \textit{TPA} module employs mutual information estimation (MIE) theory in two aspects:
(a) Minimize the upper bound of dependencies between knowledge-reflected and transition-aware features.
(b) Maximize the lower bound of consistency between distilled transitional patterns and collaborative features.
The distillation paradigm can be divided into two stages, \ie \textit{Knowledge Acquisition} and \textit{Transition Distillation}.
The first stage emphasizes the independent training of \textit{K-MLLM} to construct an initial multimodal knowledge comprehension space.
The second stage connects \textit{T-MLLM} and \textit{K-MLLM}, while ensuring the preservation of knowledge-reflected features, distilling fine-grained patterns into the knowledge space.

The contributions are summarized:
(1) We introduce a novel LLM-enhanced MSBR, which stimulates LLM potentials in integrating transitional pattern extraction with multimodal learning.
(2) We develop a dual-tower MLLM structure that decouples transition-aware and knowledge-reflected features.
(3) We propose a multimodal distillation pipeline, achieving effective transfer of transitional patterns into multimodal representations.
(4) We conduct extensive experiments to demonstrate the effectiveness of \modelname.

\section{Related Work}

\nosection{Session-based Recommendation}
% 整体分类思路是
% 1、传统session-based rec
% 2、辅助信息增强的session-based rec 
% 3、多模态session-based rec (MMSBR、DIMO)
% SBR predicts the next item based on anonymous historical sessions.
%
% Prior work can be divided into three categories, \ie ID-based, content-based, and multimodal-integrated models. 
%
Early \textit{ID-based} methods adopt Markov chains \cite{rendle2010factorizing} or various deep neural networks, like RNNs \cite{li2017neural,liu2018stamp}, GNNs \cite{wu2019session, wang2021session,xia2021self,su2023enhancing}, and attention mechanisms \cite{chen2019dynamic,su2023personalized}, to extract dynamic user intents.
To alleviate data sparsity and cold start problems, \textit{content-based} SBR incorporates auxiliary information such as attributes \cite{lai2022attribute,zhang2023bi,chen2024post}, knowledge graphs \cite{chen2023knowledge,lv2025optimize}, and descriptive texts \cite{de2021transformers4rec,hou2022towards}, promoting performance to some extent.
To further enhance representations, recent works \cite{zhang2023beyond,zhang2024disentangling} leverage multimodal learning to exploit multifaceted session patterns.
MSBR \cite{zhang2023beyond} applies a hierarchical transformer to fuse modalities, while DIMO \cite{zhang2024disentangling} disentangles modality effects in a semantic space.
However, compared with LLM-based methods, these studies demonstrate relatively constrained capability in complex semantic modeling.

\nosection{LLMs for Session-based Recommendation}
% 这块先写LLM在rec这个大领域下的现有工作的大分类，这个分类可以用prompt式和finetune式来分类，也就是一个是冻住llm，一个是微调llm。
% 然后介绍llm4sbr这个细分方向的现有工作情况，分类方式是llm-centric和llm-enhanced，你会发现这两类都是微调的，那为什么在上面那块llm4rec不采用这个分类方式呢，主要是为了差异化，其实两种分类大家都是认的，言之有理即可
% 第二块我在写intro的时候顺便全局搜过了，分成两类列在下面，你可以再搜一下我有没有漏的。（以下论文的citation已经添加进bib文件了）
% LLM-centric 
% 1、Integrating Large Language Models with Graphical Session-Based Recommendation
% 2、Re2LLM: Reflective Reinforcement Large Language Model for Session-based Recommendation
% 3、Large Language Models for Intent-Driven Session Recommendations
%
% LLM-enhanced
% 1、LLM4SBR: A Lightweight and Effective Framework for Integrating Large Language Models in Session-based Recommendation
% 2、LLM-BRec: Personalizing Session-based Social Recommendation with LLM-BERT Fusion Framework
LLMs exhibit promising potential in developing recommenders \cite{zhao2023recommender,lin2023can,lv2025collaboration} with sophisticated reasoning abilities and extensive knowledge.
LLM-based recommendation falls into two types \cite{wu2024survey}, \ie prompt-based and fine-tuning-based.
Prompt-based methods \cite{liu2024llm,hou2024large,liu2023chatgpt,gao2023chat} employs in-context learning or prompt optimization to engage LLMs as recommenders without re-training.
Fine-tuning-based methods \cite{hu2024enhancing,li2023e4srec,li2023text,bao2023bi,zhang2024notellm} update LLM parameters by training with domain-specific knowledge, making them suitable for particular tasks.
% 直接推荐：LLARA: LLaRA-Large Language-Recommendation Assistant
%TALLREC:TALLRec An Effective and Efficient Tuning Framework to Align Large Language Model with Recommendation;
%E4:E4SRec-An Elegant Effective Efficient Extensible Solution of Large Language Models for Sequential Recommendation
%A Practice-Friendly LLM-Enhanced Paradigm with Preference Parsing for Sequential Recommendation CANNOTFIND
%TMF:Triple Modality Fusion Aligning Visual, Textual, and Graph Data with Large Language Models for Multi-Behavior Recommendations
%
% 嵌入增强：
%NOTELLM:NoteLLM-A Retrievable Large Language Model for Note Recommendation; 
%ZHOUJUN:Enhancing Sequential Recommendation via LLM-based Semantic
For \textbf{LLM-based SBR}, there are two paradigms, \ie LLM-centric and LLM-enhanced.
\textit{LLM-centric SBR} \cite{LLMGR,Re2LLM,sun2024large} positions LLMs as the core recommenders, utilizing either prompts or fine-tuning for next-item predictions.
For instance, LLMGR \cite{LLMGR} employs LLMs to align textual and graph-structured data.
Re2LLM \cite{Re2LLM} develops a reflective model to utilize reinforcement learning for enhancing LLMs.
%
% However, these models suffer from inefficiency, as long prompts inherently result in slower inference speeds.
However, due to the slow inference resulted by long prompts, these models exhibit severe inefficiency and fail to realize all-ranking tasks.
%
% Besides, considering the single session as prompt confines the modeling perspective to local sequential dependencies rather than global collaborations.
%
In contrast, \textit{LLM-enhanced SBR} \cite{qiao2024llm4sbr,LLMBREC} complements SBR with semantic comprehension on representation learning.
For example, LLM4SBR \cite{qiao2024llm4sbr} introduces a framework leveraging LLMs for intent inference.
LLM-BRec \cite{LLMBREC} utilizes LLMs to generate user profiles for preference modeling.
Considering the low efficiency and adaptability of LLM-centric pipelines, we further develop LLM-enhanced SBR.
Compared with previous studies that sorely deploy LLM textual processing proficiency, our framework harnesses LLMs to exploit comprehensive session patterns under multimodal forms, addressing a critical leap in this domain.

\nosection{Multimodal Large Language Models}
% 这小节分为两块：1、介绍一些MLLM工作（可以参考bunble和molar 这两篇的related work里面的一个小节） 2、介绍现在市面上有的MLLM-based rec
% 需要进行一个详尽的搜索，现在市面上总共有哪些MLLM-based rec，以下是我找到的，可以再补充：
% 1、NoteLLM2
% 2、Harnessing Multimodal Large Language Models for Multimodal Sequential Recommendation （AAAI 2025）
% 3、Molar: Multimodal LLMs with Collaborative Filtering Alignment for Enhanced Sequential Recommendation (ACL 2025 submit)
% 4、Triple Modality Fusion: Aligning Visual, Textual, and Graph Data with Large Language Models for Multi-Behavior Recommendations
% 5、Towards unified multi-modal personalization: Large vision-language models for generative recommendation and beyond
% 6、InteraRec: Interactive Recommendations Using Multimodal Large Language Models.（pakdd 2024）
% 7、ATFLRec: A Multimodal Recommender System with Audio-Text Fusion and Low-Rank Adaptation via Instruction-Tuned Large Language Model. （arxiv 2024）
%8.MMREC- LLM Based Multi-Modal Recommender System
%9.A Multimodal Ecological Civilization Pattern Recommendation Method Based on Large Language Models and Knowledge Graph
% 
% Traditional LLMs excel in processing textual inputs but face limitations when dealing with multimodal data.
%
Multimodal large language models (MLLM) \cite{zhang2024mm,hu2024bliva,liu2024towards,luo2024deem} emerge to enable richer contextual cognition with various modalities.
Early research \cite{liu2024visual,wang2024lavin} combines separate modal-specific encoders and aligns through joint spaces.
Recent studies \cite{li2023blip,guo2023images,liu2025continual} integrate modality-specific encoders using cross-modal attention.
%
% MLLM are applied in diverse fields, including robotics \cite{driess2023palm}, recommenders \cite{yu2023multimodal,qin2024atflrec,karra2024interarec,liu2025fine}.
%
% 我觉得这块就是3句话：现有的MLLM 4 REC 同样有两种构建方式:(i) 其一\cite{}是在原有的LLM-centric rec中在intruction中增加了多模态信息，由此促进llm对于物品协同关系的全面/深入理解。(ii) 其二\cite{}是利用MLLM框架增强推荐中的多模态表征建模阶段，赋能下游推荐系统。但是，现有这些工作均无法很好适配到SBR中，因为xxxxx。
%
Research on MLLM-based recommender \cite{yu2023multimodal,qin2024atflrec,karra2024interarec,liu2025fine} remains in its nascent stages, and there are two construction pipelines.
% Significant strides have been made in advancing research within the field of Multimodal-LLM-based recommendation (MLLM4REC) \cite{ma2024triple,zhang2024notellm-2}. Broadly, MLLM4REC frameworks have two main construction pipelines.
%
\textit{The first line} \cite{ye2024harnessing,ma2024triple} integrates multimodal knowledge to enhance LLM-centric recommenders, promoting user interest modeling for prediction.
\textit{The second line} \cite{zhang2024notellm-2,luo2024molar,tian2024mmrec} employs MLLM to augment multimodal learning, providing informative embeddings for downstream tasks.
However, when adapting MLLM into SBR, studies in the first line suffer from inference computational costs while the second line, such as NoteLLM-2 \cite{zhang2024notellm-2} and Molar \cite{luo2024molar}, primarily optimize the exploration of item inherent features and overlooks critical multi-order collaborations.
% Primarily due to discrepancies in representation distributions between low-order and high-order item relationships, existing methods face notable challenges in effectively adapting to SBR scenarios.
%
% Previous studies have primarily leveraged MLLM to encode and interpret knowledge-reflective features, particularly within the visual modality of items, while largely overlooking critical interaction information crucial for SBR. 
%
% It also employs a Multi-Order Pattern Alignment framework to seamlessly infuse collaborative features into the MLLM dealing with knowledge-reflective features.
%
Contrastingly, \RAMP~builds parallel MLLMs to capture knowledge-reflected and transition-aware features, and distill transitional patterns into multimodal representations.

\section{METHODOLOGY}
% In this section, we illustrate the problem formulation of MSBR and elaborate on the proposed \modelname~framework.

\subsection{Problem Formulation}
Following previous work \cite{zhang2023beyond, zhang2024disentangling}, we formulate a typical MSBR scenario.
%
% The MSBR models takes a set of anonymous sessions $\mathcal{S}=\{S_1,S_2,\ldots,S_{|\mathcal{S}|}\}$ over the item set $\mathcal{V}=v_1,v_2,\ldots,v_{|\mathcal{V}|}\}$ as input, and 
Formally, let $\mathcal{V}=\{v_1,v_2,\ldots,v_{|\mathcal{V}|}\}$ represent all unique items in the dataset, and an anonymous session $S_i=\{v_1,v_2,\ldots,v_n\}$ contains interacted items in the chronological order where $n$ is the session length.
Each item is presented as $v_i=\{v^{id}_i,v^{mo}_i\}$, with $v^{id}$ showing its ID and $v^{mo}=\{v^{txt},v^{img}\}$ denoting multimodal information, \ie visual and textual features.
In this paper, we consider $v^{txt}$ as item descriptive text which includes title and brand, and $v^{img}$ as item image.
MSBR models aim to recommend top-$K$ items from $\mathcal{V}$ that are most likely to be interacted next.
The objective of our model is leveraging LLM to promote the multimodal session learning stage, thus improving the downstream MSBR performance.

\subsection{An overview of \modelname}
\begin{figure*}[t]
\centering
\includegraphics[width=1.0\linewidth]{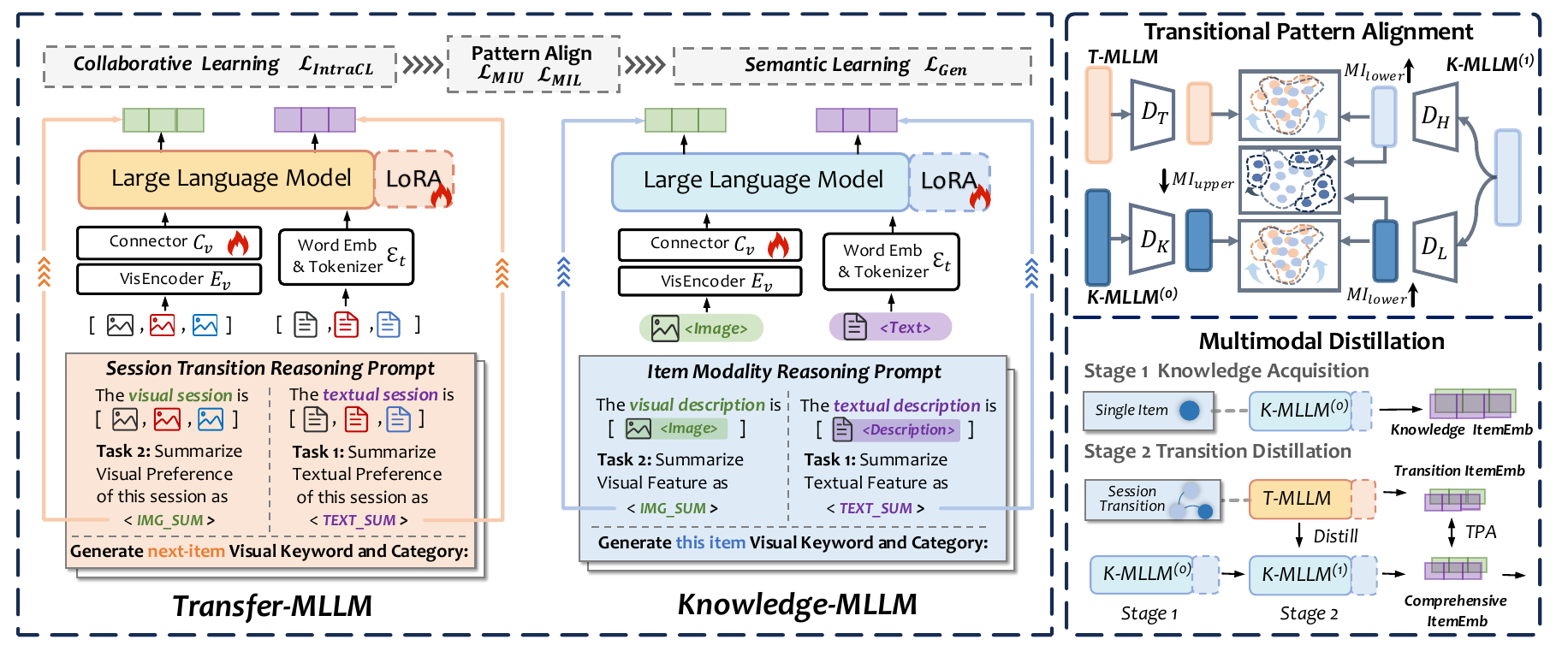}
\vspace{-0.15 in}
\caption{The framework of \modelname. (a) \modelname~consists three modules: T-MLLM extracts transition-aware features, K-MLLM depicts knowledge-reflected features, and TPA aligns transitional patterns. (b) \modelname~is optimized by multimodal distillation that first achieves knowledge acquiring and then complete transition distilling.} 
\label{framework}
% \vspace{-0.1 in}
\end{figure*}

% \nosection{Framework Overview}
\modelname~reforms LLM-enhanced MSBR with a pattern distillation paradigm, to drive LLMs to adaptively integrate transitional patterns into modal embeddings.
The framework is presented in Figure. \ref{framework}, which consists of three modules: (i) Knowledge-MLLM (K-MLLM) serves as the generator for multimodal representations, utilizing attribute inference prompts to bestow on LLM compressing proficiency of \textit{knowledge-reflected features}.
(ii) Transfer-MLLM (T-MLLM) interprets multi-order transitions with recursive reasoning instructions, decoupling \textit{transition-aware features}.
(iii) Transitional Pattern Alignment (TPA) bridges K-MLLM and T-MLLM, utilizing mutual influence estimation to resolve distribution discrepancy among sequential patterns.
\textbf{From the pipeline level}, the paradigm is divided into: \textit{Knowledge Acquisition Stage} and  \textit{Transition Distillation Stages}.
The first stage depicts item intrinsic features with dependently training K-MLLM.
The second stage extracts multi-order transition with T-MLLM and distills transition-aware semantic understanding into K-MLLM with progressive alignment.

\subsection{Knowledge Acquisition Stage}

%%% 一小段话写清楚这个模块的motivation、purpose，写这段话的时候，联合你下面每个part的标题来写
The primary item representation should be conveyed by the modality-specific information itself.
Thus the knowledge acquisition stage is essential to grant the LLM the ability to deconstruct complex intrinsic item features from diverse modality knowledge.
%

%% 讲 K-MLLM 的 prompt
\nosection{Item Modality Reasoning Prompt}
% 用一句话描述这个part的作用/亮点，也可以讲这个part分成哪几步。
% 技术细节（语言可以不高级，但不要有错误，严谨）：比较流水账的，摆公式，重要的公式后面可以有一两句话来解释。
To retrieve latent embeddings with rich semantics, we build a Knowledge-MLLM and employ a \textit{item modality reasoning prompt} to guide it in the generation task.
%
% prepare the data: text and img emb
Firstly, we collect raw textual and visual information for the item's hybrid tokenization.
For each item $v_i$, we take the brief descriptive text $\boldsymbol{v}_i^{txt}$, containing item title and brand.
Then we apply natural LLM processing to transform the textual content into embeddings:
\begin{equation}
\begin{split}
    \boldsymbol{v}_i^{txt} = [wd_{1},wd_{2},\ldots,wd_{n}],\\
    \boldsymbol{e}_i^{txt} = \boldsymbol{\mathcal{E}_t}\left(\operatorname{Tokenizer}[\boldsymbol{v}_i^{txt}]\right).
\end{split}
\end{equation}
Here, $wd_i$ is each word in text descriptions, $\boldsymbol{\mathcal{E}_t}$ indicates the word embedding table.
Recognizing advancement of foundation models, we employ the pre-trained BLIP2 \cite{li2023blip} to encode the image $\boldsymbol{v}_i^{img}$
\begin{equation}
    \boldsymbol{p}_i^{img} = \boldsymbol{E_{v}}(\boldsymbol{v}_i^{img};\Theta_v),\  \boldsymbol{e}_i^{img} = \boldsymbol{C_{v}}(\boldsymbol{p}_i^{img}),
\end{equation}
where $\boldsymbol{E_{v}}$ denotes the pre-trained vision encoder and is equipped with dataset-specific parameters $\Theta_v$.
To eliminate discrepancies of modality representation spaces \cite{zhang2024notellm}, we apply the connector $\boldsymbol{C_{v}}$ to convert visual embedding $\boldsymbol{p}_i^{img}$ into LLM word embedding space.

The design concept of the hybrid prompt lies in twofold: (i) We formulate multiple preference compression tasks to cover both visual and textual semantic learning.
(ii) The condensed special token is utilized as a trigger to fetch and optimize implicit modality interpreting from LLM.
Specifically, the prompt is illustrated as:

% \begin{mdframed}[linewidth=0.5pt, backgroundcolor=blue!10]
% A user bought an item in the \textbf{\textit{Cellphone and Accessories}} domain.
% %
% The textual description is \textbf{[$v_i^{txt}$]}, summarize the textual preference: \textbf{<TEXT\_SUM>}.
% %
% The visual description is \textbf{[$v_i^{img}$]}, summarize the visual preference: \textbf{<IMG\_SUM>}.
% %

% According to the preference summary, output two textual categories and a two-word visual keyword of this item.
% \end{mdframed}

\begin{tcolorbox}[colframe=gray!30, colback= gray!5, coltitle=black!100, title=\textit{\textbf{Item Modality Reasoning Prompt}}]
\textbf{<Instruction>:} A user bought an item in the \textbf{\textit{Cellphone and Accessories}} domain.
The visual description is \textcolor{green!50!black}{\textbf{[$v^{img}$]}}, summarize the user's visual preference: \textcolor{green!50!black}{\textbf{<IMG\_SUM>}}.
The textual description is \textcolor{violet!90}{\textbf{[$v^{txt}$]}}, summarize the user's textual preference: \textcolor{violet!90}{\textbf{<TEXT\_SUM>}}.
According to these preference summaries, output a two-word visual keyword and two textual categories of this item:

\textbf{<Answer Format>:}

\textcolor{green!50!black}{\textbf{Visual keyword:}} Screen Protector

\textcolor{violet!90}{\textbf{Categories:}} Accessory Kits, Cables

\end{tcolorbox}

In this template, we set the inference about visual and textual attributes of a single item as the \textit{primary generation task}, while taking summarization on modal-specific preference as an \textit{auxiliary learning task}.
Following \cite{wang2023label,zhang2024notellm}, we adopt a multimodal in-context-learning where <IMG\_SUM> and <TEXT\_SUM> are special tokens to retrieve latent modal embeddings construed by LLM.
%
% Intuitively,the previous hidden state of the token <TEXT\_SUM> and <IMG\_SUM> are taken as the \textbf{zero-order} textual embedding $\boldsymbol{e}^0_t$ and visual embedding $\boldsymbol{e}^0_v$.

\nosection{Multimodal Supervised Fine-tuning}
To possess balanced comprehension across modalities, we disentangle the generation task into two aspects, \ie visual keyword and category matching.
The ground truth of visual keywords is prompting results from the advanced multimodal LLM QWen-VL-Max \cite{bai2023qwen}, while item categories are from original datasets.
Typically, we reorganize pair-wise data $\mathcal{D}=\{(x_i,y_i)\}_{i=1,\ldots,|\mathcal{V}|}$, where $x_i$ and $y_i$ denote instructions and true answers, respectively.
%
% The instruction tuning of LLM follows the autoregressive objective \cite{touvron2023llama} as
% \begin{equation}
%     \max _{\Phi} \sum_{(x, y) \in \mathcal{D}} \sum_{\tau\in|y|} \log p\left(y_{\tau} \mid x, y_{<\tau}; {\Phi}\right),
% \end{equation}
% where $y_{\tau}$ refers to the $\tau$-th token of $y$ and $\Phi$ represents the parameters of the LLM.
%
To facilitate efficient optimization, we adopt Parameter Efficient Fine-Tuning (PEFT) to reduce computational cost and meanwhile achieve comparable performance.
Specifically, we choose LoRA \cite{hu2021lora} which freezes LLM weights and injects trainable low-rank decomposition matrices into each layer.
Then the optimization objective is reformed as:
\begin{equation}
\label{loss1}
    \max _{\Theta_{\mathrm{LR}}} \sum_{(x, y) \in \mathcal{D}} \sum_{\tau\in|y|} \log p\left(y_{\tau} \mid x, y_{<\tau}; \Phi, \Delta \Phi\left(\Theta_{\mathrm{LR}}\right)\right),
\end{equation}
where $y_{\tau}$ refers to the $\tau$-th token of $y$, $\Phi$ and $\Delta \Phi\left(\Theta_{\mathrm{LR}}\right)$ are the frozen LLM parameters and trainable LoRA parameters respectively.
After training to convergence, we denote the final state of Knowledge-MLLM in this stage as $K$-$MLLM^{(0)}$, which reveals the initial understanding of LLM on item knowledge-reflected features.
Correspondingly, we fetch the previous hidden state of the token <IMG\_SUM> and  <TEXT\_SUM> as the \textbf{zero-order} visual embedding $\boldsymbol{e}^0_v$ and textual embedding $\boldsymbol{e}^0_t$.

\subsection{Transition Distillation Stage}
% The evolving user intents hidden underneath sessions are decided by both item inherent attributes and collaborative dependencies.
%
To incorporate transitional pattern semantics into item multimodal representations, we design the Transition Distillation stage, which (i) extracts multi-order sequential patterns and (ii) enhances the intra-order synergy with target-aware contrastive learning.

\nosection{Session Transition Reasoning Prompt}
% 如何划分 multi-order sessions
To decouple collaborative signals at hierarchical levels, we gather multi-order meta pairs from training sessions and train a Transfer-MLLM in this stage.

\begin{definition}[Multi-order Meta Pairs]
Suppose the session as $S_i=\{v_1,v_2,\ldots,v_{n-1},v_n\}$, the $\kappa$-order meta pair set is formulated
\begin{eqnarray*}
      \delta_{\kappa}=\{[v_{m-\kappa},\ldots,v_{m-1}]\rightarrow[v_m]\  |\  1 \leq \kappa \leq n-1, 1 \leq m \leq n\}, 
\end{eqnarray*}
where $[v_{m-\kappa},\ldots,v_{m-1}]$ is meta query, $[v_m]$ is meta target.
Typically, the first-order pair is $\delta_1=\{[v_{m-1}]\rightarrow[v_m]\  |\  2 \leq m \leq n\}$ and the second-order pair is $\delta_2=\{[v_{m-2},v_{m-1}]\rightarrow[v_m]\  |\  3 \leq m \leq n\}$.
%
% $\kappa$-order pair set of target item $v_m$ is $\delta^m_{\kappa} = \{[v_{m-\kappa},\ldots,v_{m-1}]\rightarrow[v_m]\}$.
\end{definition}
We apply the sliding window mechanism on all sessions to categorize first- and second-order meta pair sets, \ie $\delta_1$ and $\delta_2$.
Then we propose a \textit{session transition reasoning prompt} as follows:

\begin{tcolorbox}[colframe=gray!30, colback= gray!5, coltitle=black!100, title=\textit{\textbf{Session Transition Reasoning Prompt}}]
\textbf{<Instruction>:} A user has bought several items successively in the \textbf{\textit{Cellphone and Accessories}} domain.

The visual description of the session is \{ Item1: \textcolor{green!50!black}{\textbf{[$v^{img}_{m-2}$]}}, Item2: \textcolor{green!50!black}{\textbf{[$v^{img}_{m-1}$]}} \}, summarize the user's visual preference: \textcolor{green!50!black}{\textbf{<IMG\_SUM>}}.
The textual description of the session is \{ Item1: \textcolor{violet!90}{\textbf{[$v^{txt}_{m-2}$]}}, Item2: \textcolor{violet!90}{\textbf{[$v^{txt}_{m-1}$]}} \}, summarize the user's textual preference: \textcolor{violet!90}{\textbf{<TEXT\_SUM>}}.

Predict the next item of this session, output a two-word visual keyword and two textual categories of this item:

\textbf{<Answer Format>:}

\textcolor{green!50!black}{\textbf{Visual keyword:}} Screen Protector

\textcolor{violet!90}{\textbf{Categories:}} Accessory Kits, Cables

\end{tcolorbox}
Note that $T$-$MLLM$ is trained through several iterations with multi-order meta pairs.
For the $\kappa$-th iteration, we provide modal information of $\kappa$-order meta query, and expect it to infer the user's potential interests and make predictions on the meta target $v_m$.
% This session prompt is also a combination of multimodal in-context-learning tasks and language generation tasks.
%
% For \textit{Item1}, we insert the raw text description of the \textit{first} item in the meta query from $\kappa$-order pairs.
% %
% For \textit{Item2}, with special token <IMG\_SUM> and <TEXT\_SUM>, we inject the latent summary embedding of the ($\kappa$-1)-order pair from the previous state $T\text{-}MLLM^{(\kappa-1)}$:
% \begin{equation}
%     [\boldsymbol{c}^{\kappa-1}_{v,m}, \boldsymbol{c}^{\kappa-1}_{t,m}] \leftarrow T\text{-}MLLM^{(\kappa-1)}(\delta^m_{\kappa-1}).
% \end{equation}
% Thus, \textit{Item2} represents an abstract item, containing lower-order transitional patterns.
% %
% Note that $\boldsymbol{c}^{\kappa-1}_{v,m}$ and $\boldsymbol{c}^{\kappa-1}_{t,m}$ are saved with dimension reduction and transformed into LLM space with projectors $p_v$ and $p_t$.
% %
% By integrating low-order cognition into the high-level reasoning process, we refine transitional pattern modeling and also promote inference efficiency.
%
% Compared to treating the entire session as input, we significantly reduces the prompt length and promotes inference efficiency.

\nosection{Intra-order Contrastive Learning}
% We also utilize LoRA to fintune $T$-$MLLM$.
% 为什么要做intra-order对比学习的motivation
% 多层次学习类似于多级消息的传递容易导致smoothing 甚至是表征塌缩，因此1、在一个order 不同物品的pattern之间应该是希望具有可辨识度discriminative
% 2、一个order 某个物品对应的pattern是diverse的，希望具有一致性 uniformity
Intuitively, multi-order transition learning, akin to hierarchical message passing, is prone to smoothing effects and potential distribution collapse.
Thus, we strengthen intra-order learning by (a) increasing the discernibility of intra-order patterns among different items.
(b) ensuring the uniformity of various intra-order patterns from the same item.

Specically, we classify $\kappa$-order meta pairs according to their meta target.
The meta pairs with the same item target are regarded as positive pairs since they exhibit similar user preference on the current sessions, and otherwise negative pairs.
With special token <IMG\_SUM> and <TEXT\_SUM>, we retrieve latent summary embedding of the $\kappa$-order pair for the meta target $m$ as:
\begin{equation}
    [\boldsymbol{c}^{\kappa}_{v,m}, \boldsymbol{c}^{\kappa}_{t,m}] \leftarrow T\text{-}MLLM(\delta^m_{\kappa}).
\end{equation}
Then we employ intra-order contrastive learning to facilitate discriminative and uniform multi-order pattern exploration:
\begin{equation}
\small
\begin{aligned}
    \mathcal{L}^o_{IntraCL} = -\frac{1}{2B}&\sum^{2B}_{m=1}\log \frac{\exp(\text{sim}(\boldsymbol{c}^{\kappa}_{o,m}, \boldsymbol{c}^{\kappa+}_{o,m}) / \tau)}{\sum_{k\in[2B] \textbackslash \{m\}} \exp(\text{sim}(\boldsymbol{c}^{\kappa}_{o,m}, \boldsymbol{c}^{\kappa}_{o,k}) / \tau)}, o\in\{v,t\}\\
    &\mathcal{L}_{IntraCL} = \mathcal{L}^v_{IntraCL} + \mathcal{L}^t_{IntraCL}.
\end{aligned}
\nonumber
\end{equation}
Here, $2B$ denotes the size of minibatch, and positive meta pairs $(\boldsymbol{c}^{\kappa}_{o,m}, \boldsymbol{c}^{\kappa+}_{o,m})$ are latent summary embedding of meta queries from $\delta^m_{\kappa}$, sharing the same meta target $m$.
We utilize the cosine similarity here.
Intra-order contrastive learning is independently applied on two modalities to avoid introducing inter-modal distribution bias.

\subsection{Transitional Pattern Alignment}
% K-MLLM depicts knowledge-reflected features while T-MLLM extracts transition-aware features.
%
To enable K-MLLM producing generalizable multimodal embeddings, we propose a Transitional Pattern Alignment (TPA) accompanying the transition distillation stage.
The alignment aims to (i) transfer context-aware sequential dependencies into static item multimodal representations, and (ii) enhance a balanced LLM cognition on transition-aware and knowledge-reflected features.

\nosection{Dual-level Feature Disentanglement}
Considering item embeddings as a unification of \textit{knowledge-reflected} and \textit{transition-aware} features, we first decompose this dual-level impact.
In the distillation stage, we initialize K-MLLM with $K$-$MLLM^{(0)}$ and retrieve item visual and textual item summary embeddings, \ie $\boldsymbol{e}^{\kappa}_{v}$ and $\boldsymbol{e}^{\kappa}_{t}$.
We first project item multimodal summary embeddings into two independent feature spaces:
\begin{equation}
\label{decouple}
    \boldsymbol{z}^{\kappa}_{o} = D^o_{H}(\boldsymbol{e}^{\kappa}_{o}), \ \boldsymbol{w}^{\kappa}_{o} = D^o_{L}(\boldsymbol{e}^{\kappa}_{o}), \ o\in\{v,t\},
\end{equation}
where $D^o_{H}$ and $D^o_{L}$ are modal-specific multi-layer perceptron.
$D^o_{H}$ serves as the transition-aware feature decoder while $D^o_{L}$ as the knowledge-reflected feature decoder.
Then we propose a Hierarchical Mutual Information Estimator (HMIE) to achieve disentanglement.
%
% Formally, mutual information measures the dependence between two variables, the MI between $\boldsymbol{z}^{\kappa}_{o}$ and $\boldsymbol{w}^{\kappa}_{o}$ is defined as:
% \begin{equation}
%     \mathcal{I}(\boldsymbol{z}^{\kappa}_{o} ; \boldsymbol{w}^{\kappa}_{o})=\mathbb{E}_{p(\boldsymbol{z}^{\kappa}_{o}, \boldsymbol{w}^{\kappa}_{o})}\left[\log \frac{p(\boldsymbol{z}^{\kappa}_{o}, \boldsymbol{w}^{\kappa}_{o})}{p(\boldsymbol{z}^{\kappa}_{o}) p(\boldsymbol{w}^{\kappa}_{o})}\right] .
% \end{equation}
We utilize a variational distribution $q_{\theta}(\boldsymbol{w}^{\kappa}_{o}|\boldsymbol{z}^{\kappa}_{o})$ to approximate the conditional distribution $p(\boldsymbol{w}^{\kappa}_{o}|\boldsymbol{z}^{\kappa}_{o})$, then the MI can be constrained by a variational contrastive log-ratio upper bound \cite{cheng2020club}:
\begin{equation}
\begin{aligned}
\mathcal{I}_{\mathrm{CLUB}}(\boldsymbol{z}^{\kappa}_{o} ; \boldsymbol{w}^{\kappa}_{o})=  &\mathbb{E}_{p(\boldsymbol{z}^{\kappa}_{o}, \boldsymbol{w}^{\kappa}_{o})}\left[\log q_\theta(\boldsymbol{w}^{\kappa}_{o} \mid \boldsymbol{z}^{\kappa}_{o})\right] \\
 &-\mathbb{E}_{p(\boldsymbol{z}^{\kappa}_{o})} \mathbb{E}_{p(\boldsymbol{w}^{\kappa}_{o})}\left[\log q_\theta(\boldsymbol{w}^{\kappa}_{o} \mid \boldsymbol{z}^{\kappa}_{o})\right].
\end{aligned}
\end{equation}
With $N$ sample pairs $\{(\boldsymbol{z}^{\kappa}_{o,i}, \boldsymbol{w}^{\kappa}_{o,i})\}_{i=1}^N$, we apply unbiased estimation:
\begin{equation}
\small
    \begin{aligned}
& \mathcal{L}_{MIU}=\widetilde{\mathcal{I}}_{\mathrm{CLUB}}=\frac{1}{N} \sum_{i=1}^N[\log q_\theta(\boldsymbol{w}^{\kappa}_{o,i} \mid \boldsymbol{z}^{\kappa}_{o,i})-\frac{1}{N} \sum_{j=1}^N \log q_\theta(\boldsymbol{w}^{\kappa}_{o,j} \mid \boldsymbol{z}^{\kappa}_{o,i})].
% =\frac{1}{N^2} \sum_{i=1}^N \sum_{j=1}^N\left[\log q_\theta\left(\boldsymbol{y}_i \mid \boldsymbol{x}_i\right)-\log q_\theta\left(\boldsymbol{y}_j \mid \boldsymbol{x}_i\right)\right] \\
\end{aligned}
\nonumber
\end{equation}
Here, $\log q_\theta(\boldsymbol{w}^{\kappa}_{o,i} \mid \boldsymbol{z}^{\kappa}_{o,i})$ represents the conditional log-likelihood of positive sample pairs while $\log q_\theta(\boldsymbol{w}^{\kappa}_{o,j} \mid \boldsymbol{z}^{\kappa}_{o,i})$ provides the negative ones.
Through minimizing $\mathcal{L}_{MIU}$, we disentangle collaborative effects from item intrinsic features.

\nosection{Distilled Pattern Alignment}
To transfer informative session dependencies, we simultaneously (i) Align decoupled transition-aware features from $K$-$MLLM$ with transitional patterns extracted from $T$-$MLLM$.
(ii) Align decoupled knowledge-reflected features from $K$-$MLLM^{(1)}$ with $K$-$MLLM^{(0)}$ to prevent knowledge forgetting.
From $T$-$MLLM$, we collect the $\kappa$-order meta query embeddings $\boldsymbol{c}^{\kappa*}_{t,m}$ and $\boldsymbol{c}^{\kappa*}_{v,m}$ of target item $m$:
% \ie, $\boldsymbol{c}^{\kappa}_{t,m}$ and $\boldsymbol{c}^{\kappa}_{v,m}$.
\begin{equation}
    \overline{\boldsymbol{c}}^{\kappa}_{t,m}=AvgPool(\boldsymbol{c}^{\kappa*}_{t,m}), \  \overline{\boldsymbol{c}}^{\kappa}_{v,m}=AvgPool(\boldsymbol{c}^{\kappa*}_{v,m}),
\end{equation}
where the average pooling is conducted to obtain the general pattern of $\kappa$-order transitions from the session context.
Similarly, we encode them into latent feature space $\boldsymbol{r}^{\kappa}_{o} = D^o_{T}(\overline{\boldsymbol{c}}^{\kappa}_{o}),\  o \in \{v,t\}$.
% \begin{equation}
%     \boldsymbol{r}^{\kappa}_{o} = D^o_{T}(\overline{\boldsymbol{c}}^{\kappa}_{o}),\  o \in \{t,v\}.
% \end{equation}
Then different from HMIE, we propose an Aligning Mutual Information Estimator (AMIE) here to distill high-order signals $\boldsymbol{r}^{\kappa}_{o}$ into the collaborative part $\boldsymbol{z}^{\kappa}_{o}$ of modal representations.
Mutual information is treated as the KL-divergence between the joint and marginal distributions, which admits following dual representation \cite{belghazi2018mutual}
% \begin{equation}
%      \mathcal{I}(\boldsymbol{z}^{\kappa}_{o} ; \boldsymbol{r}^{\kappa}_{o}) = \mathrm{KL}(p(\boldsymbol{z}^{\kappa}_{o},\boldsymbol{r}^{\kappa}_{o})\|p(\boldsymbol{z}^{\kappa}_{o})p(\boldsymbol{r}^{\kappa}_{o})).
% \end{equation}
% Recall the KL-divergence admits following dual representation \cite{belghazi2018mutual}
\begin{equation}
    \mathcal{I}_{\mathrm{MINE}}(\boldsymbol{z}^{\kappa}_{o} ; \boldsymbol{r}^{\kappa}_{o})= \operatorname{sup}_{f:\Omega\rightarrow\mathbb{R}}\mathbb{E}_{p(\boldsymbol{z}^{\kappa}_{o},\boldsymbol{r}^{\kappa}_{o})}[f]-\mathrm{log}(\mathbb{E}_{p(\boldsymbol{z}^{\kappa}_{o})p(\boldsymbol{r}^{\kappa}_{o})}[e^f]),
\nonumber
\end{equation}
where the supremum is evaluated over all functions $f$.
% \begin{proof}[Proof. Mutual Information Lower Bound]
% Denote the distribution of $p(\boldsymbol{z}^{\kappa}_{o},\boldsymbol{c}^{\kappa}_{o})$ as $\mathbb{P}$ and $p(\boldsymbol{z}^{\kappa}_{o})p(\boldsymbol{c}^{\kappa}_{o})$ as $\mathbb{Q}$, introduce the Gibbs distribution $\mathbb{G}$ defined by $d\mathbb{G}=\frac{1}{Z}e^fd\mathbb{Q}$, where $Z=\mathbb{E}_{\mathbb{Q}}[e^f]$.
% %
% Then we calculate the gap:
% \begin{equation*}
% \begin{aligned}
%    &\Delta =  \mathrm{KL}(\mathbb{P}\|\mathbb{Q})-(\mathbb{E}_{\mathbb{P}}[f]-\mathrm{log}(\mathbb{E}_{\mathbb{Q}}[e^f])) \\
%    &=\mathbb{E}_{\mathbb{P}}\left[\mathrm{log}\frac{d\mathbb{P}}{d\mathbb{Q}}-\mathrm{log}\frac{d\mathbb{G}}{d\mathbb{Q}}\right]=\mathbb{E}_{\mathbb{P}}\mathrm{log}\frac{d\mathbb{P}}{d\mathbb{G}}=\mathrm{KL}(\mathbb{P}\|\mathbb{G}) \geq 0.
% \end{aligned}
% \end{equation*}
% \end{proof}
% Considering the positivity of $\mathrm{KL}(\mathbb{P}\|\mathbb{G})$, 
Then we have the lower bound of mutual information:
\begin{equation}
    \mathcal{L}^h_{MIL}= \mathbb{E}_{p(\boldsymbol{z}^{\kappa}_{o}, \boldsymbol{r}^{\kappa}_{o})}[f_{\phi}(\boldsymbol{z}^{\kappa}_{o}, \boldsymbol{r}^{\kappa}_{o})]-\log (\mathbb{E}_{p(\boldsymbol{z}^{\kappa}_{o}) p(\boldsymbol{r}^{\kappa}_{o})}[e^{f_{\phi}(\boldsymbol{z}^{\kappa}_{o}, \boldsymbol{r}^{\kappa}_{o})}]),
\nonumber
\end{equation}
where $f_{\phi}$ is the score function approximated by multilayer perception.
Meanwhile, we input the zero-order of visual embeddings $\boldsymbol{e}^0_{v}$ and textual $\boldsymbol{e}^0_{t}$ from $K$-$MLLM^{(0)}$ into the projector $\overline{\boldsymbol{e}}_{o} = D^o_{K}(\boldsymbol{e}^{0}_{o}),\  o \in \{v,t\}$.
% \begin{equation}
%     \overline{\boldsymbol{e}}_{o} = D^o_{K}(\boldsymbol{e}^{0}_{o}),\  o \in \{t,v\}.
% \end{equation}
To preserve knowledge-reflected features, we align the decoupled inherent embeddings in Eq.(\ref{decouple}) with zero-order embeddings:
\begin{equation}
    \mathcal{L}^s_{MIL}= \mathbb{E}_{p(\boldsymbol{w}^{\kappa}_{o},  \overline{\boldsymbol{e}}^{\kappa}_{o})}[f_{\phi}(\boldsymbol{w}^{\kappa}_{o},  \overline{\boldsymbol{e}}^{\kappa}_{o})]-\log (\mathbb{E}_{p(\boldsymbol{w}^{\kappa}_{o}) p( \overline{\boldsymbol{e}}^{\kappa}_{o})}[e^{f_{\phi}(\boldsymbol{w}^{\kappa}_{o},  \overline{\boldsymbol{e}}^{\kappa}_{o})}]).
\nonumber
\end{equation}

By maximizing the lower bound of mutual information between the cognition from each state of $K$-$MLLM$ and $T$-$MLLM$ and concurrently enhancing comprehension of self-knowledge, we effectively align and incorporate distilled transitional patterns.

\subsection{Multimodal Distillation Optimization}

\nosection{Distillation Optimization}
The overall training of two MLLMs can be divided into multi-stages.
For the Knowledge Acquisition Stage, we train $K$-$MLLM^{(0)}$ individually, and the optimization objective is presented in Eq.(\ref{loss1}).
For the Transition Distillation Stage, we first utilize meta pair sets to instruct $T$-$MLLM$.
Denote the visual keyword and category generation loss as $\mathcal{L}_{Gen}$, the combined training objective for $T$-$MLLM$ is:
\begin{equation}
\mathcal{L}_{T-MLLM}=\mathcal{L}_{Gen}+\mu\mathcal{L}_{IntraCL}.
\end{equation} 
% \zqy{should be divided by (1+u) }
After $T$-$MLLM$ is optimized, with the inference results of latent summary embeddings on transitional patterns, we subsequently train $K$-$MLLM^{(1)}$ based on: 
\begin{equation}
\mathcal{L}_{K-MLLM}=\mathcal{L}_{Gen}+\gamma_1\mathcal{L}_{MIU}+\gamma_2\mathcal{L}^h_{MIL}+\gamma_3\mathcal{L}^s_{MIL}.
\end{equation} 
% \zqy{should be divided by (1+r1+r2)}

\nosection{Inference on Downstream MSBR}
Considering the average length of sessions, we set the highest order $\kappa=2$.
After iteratively optimizing two parallel MLLMs, we utilize the final state of $K$-$MLLM$ to perform \textit{one-time inference} on all items, and then obtain the comprehensive visual and textual summary representations as $\boldsymbol{\hat{e}}_v$ and $\boldsymbol{\hat{e}}_t$.
We feed them into a tunable adapter to further project them into recommendation space as \cite{liu2024llm}:
\begin{equation}
    \boldsymbol{e}^{rec}_o = \boldsymbol{W}^2_o(\boldsymbol{W}^1_o \boldsymbol{\hat{e}}^2_o+\boldsymbol{b}^1_o)
    +\boldsymbol{b}^2_o,\  o \in \{v,t\},
\end{equation}
where $\boldsymbol{e}^{rec}_v$ and $\boldsymbol{e}^{rec}_t$ are final multimodal embeddings adopted in downstream MSBR.
Note that \modelname~serves as a pluggable embedding model to any specific MSBR, thus no modifications are necessary for downstream model structures.
We train MSBR models with multimodal embeddings to complete recommendation tasks.

\section{Experiments and analysis}

We conduct experiments to address following questions:
\textbf{RQ1}: How does our model perform compared with SBR, MSBR, and LLM-based MSBR models?
\textbf{RQ2}: How does each designed module contribute to the performance?
\textbf{RQ3}: What are the specific properties of our model, such as \textit{generation quality}, \textit{efficiency}, \textit{distillation interpretability}, \textit{parameter sensitivity}, and \textit{scalability} under various scenarios?
%
% \textbf{RQ4}: How do we prove that the TPA module aligns transitional patterns?
% %
% \textbf{RQ5}: How hyper-parameters affects \modelname?
% %
% \textbf{RQ6}: How robust is our model across various scenarios?
%
\begin{table}
\small
  \centering
  \caption{Statistics of datasets. Here we report the number of sessions (Session), number of interactions (Interact), the average length of sessions (AvgL), and the number of first-order ($1^{st}$) and second-order ($2^{nd}$) pairs.}
     % \vspace{-0.2cm}
    \begin{tabular}{cccccccc}
    \toprule
    
    \textbf{Datasets}  & Session & Interact & AvgL & Item &   $1^{st}$-pair  &   $2^{nd}$-pair\\
    \hline
    \textbf{Cellphone} &8,983& 35,443& 3.95& 2,783 &16,364 & 11,256\\
    \textbf{Grocery} & 43,949& 177,702& 4.04& 5,462 & 48,592 & 41,960 \\
    \textbf{Sports} & 34,246& 148,250& 4.33& 6,546 & 53,416 &43,440\\

    % \#training session &54,647 &89,281 & 148,431\\
    % \#validation session &15,613 &25,509 & 42,409\\
    % \#testing session &7,766 & 12,758& 21,119\\
    
\bottomrule
    \label{tab:data}
    \end{tabular}%
    \vspace{-0.15in}
\end{table}%

\begin{table*}[t]
\centering
\caption{Experimental results on three datasets. The best results are boldfaced and the second-best results are underlined. (*) denotes a significant improvement over the best baseline (t-test p<0.05).}
% \vspace{-0.4cm}
\label{tab:compare}
\resizebox{\linewidth}{!}{
\begin{tabular}{c|cccc|cccc|cccc}
\toprule
\multirow{2.2}{*}{\textbf{Datasets}}
& \multicolumn{4}{c|}{\textbf{Cellphone \& Accessories}}
& \multicolumn{4}{c|}{\textbf{Grocery \& Food}}
& \multicolumn{4}{c}{\textbf{Sports \& Outdoors}}\\
\cmidrule(lr){2-5} \cmidrule(lr){6-9} \cmidrule(lr){10-13}
&HR@5 &NDCG@5 &HR@10 &NDCG@10 &HR@5 &NDCG@5 &HR@10 &NDCG@10 &HR@5 &NDCG@5 &HR@10 &NDCG@10\\
\midrule
\textbf{NARM} & 0.1645&	0.1522&	0.2204&	0.1668&	0.4175&	0.3911&	0.4296&	0.4038& 0.2316&	0.2301&	0.2709&	0.2352\\
\textbf{SASRec} & 0.1903&	0.1656&	0.2419&	0.1782&	0.4414&	0.4105&	0.4591&	0.4363& 0.2528&	0.2498&	0.2931&	0.2572\\
\textbf{SRGNN} & 0.1832&	0.1644&	0.2375&	0.1773&	0.4321&	0.4072&	0.4555&	0.4318&0.2415&	0.2432&	0.2806&	0.2529 \\
\textbf{MSGIFSR} & 0.1977&	0.1765&	0.2493&	0.1864&	0.4552&	0.4222&	0.4794&	0.4479& 0.2635&	0.2587&	0.3019&	0.2688\\
\textbf{AttMixer} & 0.2097&	0.1809&	0.2528&	0.1897&	0.4641&	0.4319&	0.5007&	0.4542& 0.2760&	0.2614&	0.3138&	0.2799\\

\midrule
\textbf{MGS} & 0.1918&	0.1697&	0.2342&	0.1785&	0.4532&	0.4278&	0.4835&	0.4398&	0.2608&	0.2556&	0.3019&	0.2733\\
\textbf{UniSRec} & 0.2033&	0.1785&	0.2468&	0.1876&	0.4645&	0.4324&	0.5003&	0.4531&	0.2778&	0.2616&	0.3142&	0.2805\\
\textbf{CoHHN} & 0.2108&	0.1820&	0.2536&	0.1909&	0.4705&	0.4373&	0.5141&	0.4519&	0.2827&	0.2654&	0.3197&	0.2832\\
\textbf{BiPNet} & 0.2115&	0.1823&	0.2502&	0.1892&	0.4817&	0.4432&	0.5188&	0.4557&	0.2831&	0.2651&	0.3205&	0.2813\\
\midrule

\textbf{MMSBR} & 0.2234&	0.1889&	0.2647&	0.1934&	0.5038&	0.4547&	0.5309&	0.4678 & 0.3067&	0.2790&	0.3311& 0.2865\\
\textbf{DIMO} & 0.2305&	0.1943&	0.2773&	0.2089&	0.5045&	0.4521&	0.5325&	0.4616&	0.3118&	0.2846&	0.3424&	0.2939\\

\midrule 
% BLIP-2_{+MMSBR} & \\
$\textbf{BLIP-2}_{+DIMO}$ & 0.2311&	0.1943&	0.2775&	0.1993&	0.5052&	0.4525&	0.5331&	0.4622&	0.3018&	0.2699&	0.3284&	0.2834\\
% LLaVA & \\
% Qwen2-VL_{+MMSBR} & \\
$\textbf{Qwen2-VL}_{+DIMO}$ & 0.2378	&0.1974	&0.2833	&0.2092 &0.5173	&0.4656	&0.5412	&0.4731 &0.3164	&0.2875	&0.3477	&0.2952\\
% NoteLLM2-NA_{+MMSBR} & \\
$\textbf{NoteLLM2-NE}_{+DIMO}$ & \underline{0.2397}	&\underline{0.1975}	&\underline{0.2851}	&\underline{0.2107} & \underline{0.5241}	 &0.4663	 &0.5487	 &0.4712 & 0.3172	&0.2847	&0.3503	&0.2941\\
% RAMP-NA_{+MMSBR} & \\
\modelname$\textbf{-NA}_{+DIMO}$& 0.2381	&0.1968	&0.2836	&0.2083	&0.5237	&\underline{0.4669}	&\underline{0.5505}	&\underline{0.4756}	&\underline{0.3179}	&\underline{0.2883}	&\underline{0.3512}	&\underline{0.2971}\\

\midrule
\rowcolor{gray!18} 
$\modelname_{+DIMO}$ & \textbf{0.2503*}	& \textbf{0.2078*}	&\textbf{0.2971*}	&\textbf{0.2231*}&\textbf{0.5398*}	&\textbf{0.4833*}&\textbf{0.5688*}	&\textbf{0.4879*}&\textbf{0.3294*}	&\textbf{0.2983*}	&\textbf{0.3622*}	&\textbf{0.3087*}\\

% $p$-value & $1e^{-2}$ & $7e^{-3}$ & $1e^{-2}$  & $5e^{-3}$  & $9e^{-3}$ & $6e^{-3}$ & $1e^{-2}$ & $7e^{-3}$ & $8e^{-3}$ & $1e^{-2}$ & $6e^{-3}$ & $1e^{-2}$\\

\bottomrule

\end{tabular}
%}
}
% \vspace{-0.10 in}
\end{table*}

\subsection{Experimental Setup}
\nosection{Datasets}
We evaluate our model on three datasets covering diverse domains from \textit{Amazon} platform, \ie Cellphones \& Accessories, Grocery \& Food, and Sports \& Outdoors.
As for recommender training, we follow previous MSBR \cite{zhang2023beyond,zhang2024disentangling} to collect user behaviors within one day as sessions and take the latest interacted item as the prediction target.
We filter the items that appear less than 5 times and sessions containing less than two items.
For multimodal learning, we delete items with invalid or missing descriptive text/images.
The training, validation, and testing set are split with the ratio 7:2:1.
We apply the sliding window method on training sessions and filter out the $1^{st}$- and  $2^{nd}$-order meta pairs whose meta target appears only one time, they are used to train $T$-$MLLM$.
$K$-$MLLM^{(0)}$ is trained on the whole item set, while $K$-$MLLM^{(1)}$ is trained on the meta target set of first-order ($1^{st}$) and  second-order ($2^{nd}$) pairs.
We present the detailed statistics in Table \ref{tab:data}.

\nosection{Baseline Algorithms}
We compare \textbf{four types} of baselines:

\nosection{\textit{ID-based Session-based Recommendation}}
(1) \textbf{NARM} \cite{li2017neural} utilizes GRU to model sessions.
(2) \textbf{SASRec} \cite{kang2018self} captures sequential patterns with transformers.
(3) \textbf{SRGNN} \cite{wu2019session} employs gated GNN to represent complex item transitions.
(4) \textbf{MSGIFSR} \cite{guo2022learning} utilizes heterogeneous session graphs to identify multi-granularity user intents.
(5) \textbf{AttMixer} \cite{zhang2023efficiently} proposes a multi-level attention mixture network for reasoning transitions in sessions.

\nosection{\textit{Content-based Session-based Recommendation}}
(6) \textbf{MGS} \cite{lai2022attribute} integrates item attributes to delineate intents.
(7) \textbf{UniSRec} \cite{hou2022towards} constructs universal representations with descriptive text.
(8) \textbf{CoHHN} \cite{zhang2022price} incorporates price preference into modeling.
(9) \textbf{BiPNet} \cite{zhang2023bi} uses hypergraphs to explore price and interest-based preferences.

\nosection{\textit{Multimodal Session-based Recommendation}}
(10) \textbf{MMSBR} \cite{zhang2023beyond} models descriptive and numerical information in a unified attention-based framework.
(11) \textbf{DIMO} \cite{zhang2024disentangling} disentangles ID and modality effects with multi-view self-supervision learning.

\nosection{\textit{LLM-enhanced Session-based Recommendation}}
(12) \textbf{BLIP-2} \cite{li2023blip} utilizes a lightweight querying transformer to align frozen image encoders and LLMs.
(13) \textbf{Qwen2-VL} \cite{wang2024qwen2} employs dynamic resolution for multimodal LLM.
We design specific prompts to retrieve embeddings from Qwen2-VL-7B.
(14) \textbf{NoteLLM-2} \cite{zhang2024notellm} employs multimodal compression and late fusion to enhance LLM learning.
For a fair comparison, we reproduce a version named \textbf{NoteLLM2-NE}, which uses Llama2-7B as the LLM backbone and keeps other technical details unchanged.
(15) \modelname\textbf{-NA} is the naive version of \modelname~which only realizes the first state of K-MLLM without transition distillation.

\nosection{Evaluation Protocols}
% 要说明我们的MLLM是抽取了embedding之后放到下游的MSBR上做推荐任务。
% HR NDCG 
% ACC (category vskw)
Since \modelname~serves as an embedding model, we evaluate the downstream SBR performance which uses multimodal embeddings generated by \modelname.
We choose DIMO \cite{zhang2024disentangling}, the latest MSBR SOTA, as the downstream SBR.
As for LLM-enhanced baselines, we extract their generated multimodal embeddings in the same way.
We employ commonly used ranking-based metrics, \ie HR and NDCG, where the ranked list cut-off is set 5 and 10.
We conduct all experiments five times and report the average results, and complete statistical significance tests by performing $t$-tests.
%
% Besides, to assess the training of \modelname, we propose a new metric $GR=p_{mc}/p_{all}$ for measuring the quality of generation results from $T$-$MLLM$ and $K$-$MLLM$, where $p_{mc}$ is the number of generated answers matching the ground truth and $p_{all}$ denotes the prompt amount.
% %
% Correspondingly, we have $GR_{cat}$ and $GR_{vsk}$ to represent the category and visual keyword prediction accuracy respectively.

\nosection{Implementation Details}
We select Llama2-7B \cite{touvron2023llama} as the LLM backbone.
All fine-tuning are conducted on 4 NVIDIA A40 GPUs, with a batchsize of 512.
We employ a warm-up strategy for the learning rate, initiated with 1/100 of the maximum rate, and decay it with a cosine scheduler over steps.
The dimension of LLM embeddings is set 4096, and the latent dimension of embeddings from $D^o_{H}$, $D^o_{L}$, $D^o_{T}$, and $D^o_{K}$ is 128.
Specifically, we explore the impact of loss weight $\mu$ and $\gamma_1$ by varying them in $\{0.001,0.005,0.007,0.01,0.05\}$, $\gamma_2$ and $\gamma_3$ in range $\{0.1,0.5,1.0,5.0,10.0\}$.
For the downstream MSBR, we follow settings in \cite{zhang2024disentangling} with embedding size as 100, batchsize as 50, learning rate as 0.001.
Other hyperparameters of all models are determined via grid search with validation sets.

\begin{table}[t]
\renewcommand\arraystretch{0.9}
\centering
\caption{Ablation study on key components of \modelname.}

\label{tab:ablation}
\resizebox{0.9\linewidth}{!}{
\begin{tabular}{c|cc|cc}
\toprule
\multirow{2.2}{*}{\textbf{Variants}}
& \multicolumn{2}{c|}{\textbf{Cellphone}}
& \multicolumn{2}{c}{\textbf{Grocery}}\\
\cmidrule(lr){2-3} \cmidrule(lr){4-5} 
 &HR@10 &NDCG@10 & HR@10 &NDCG@10 \\

\midrule 
\textbf{w/o IntraCL-v} & 0.2930	&0.2197		&0.5674	&0.4869\\
\textbf{w/o IntraCL-t} & 0.2908	&0.2165		&0.5633	&0.4846\\
\textbf{w/o Gen-T}& 0.2893	&0.2127		&0.5616	&0.4825\\
\midrule 

\textbf{w/o MIL-h} 	&0.2861	&0.2112		&0.5508	&0.4762\\
\textbf{w/o MIL-l} 	&0.2886	&0.2145		&0.5587	&0.4824\\
\textbf{w/o MIU} 	&0.2865	&0.2117 	&0.5519	&0.4758\\
\textbf{w/o Gen-K}	&0.2785	&0.2013		&0.5329	&0.4620\\
\midrule 

% \textit{w/o MIL-h} & 0.2419	&0.1989	&0.2861	&0.2112	&0.5241	&0.4667	&0.5508	&0.4762\\
% \textit{w/o MIL-l} & 0.2454	&0.2023	&0.2875	&0.2134	&0.5298	&0.4743	&0.5587	&0.4824\\
% \textit{w/o MIU} & 0.2425	&0.1976	&0.2865	&0.2107 &0.5263	&0.4671	&0.5519	&0.4758\\
% \textit{w/o GEN-K}& 0.2314	&0.1949	&0.2785	&0.2013	&0.5057	&0.4528	&0.5329	&0.4620\\

% \modelname & \textbf{0.2481} & \textbf{0.2052} & \textbf{0.2948} &\textbf{0.2149} &\textbf{0.5376} &\textbf{0.4817} & \textbf{0.5688} & \textbf{0.4879}\\
\rowcolor{gray!18} \modelname & \textbf{0.2971} &\textbf{0.2231} & \textbf{0.5688} & \textbf{0.4879}\\

\bottomrule

\end{tabular}
}
% \vspace{-0.15 in}
\end{table}

\subsection{Overall Performance (RQ1)}
%% 一定要分点写
%% 每一点：现象+原因推测
We evaluate the performance of \modelname~with extensive baselines and report the results in Table \ref{tab:compare}, where the following observations are noted:
1) \textit{Leveraging content information improves recommendation accuracy.}
Models that incorporate item content or modalities surpass most ID-based approaches, underscoring the efficacy of extracting content-aware associations.
%
% underscoring the advantage of utilizing modality information to effectively capture and represent users' evolving preferences.
%
2) \textit{Harnessing multimodal features has advantages over utilizing auxiliary content in the unimodal form.}
The latest MSBR SOTAs, \ie MMSBR and DIMO, perform better than content-based models, which highlights the essential of exploiting multimodal semantics to capture user evolving interest.
%
% highlighting the critical role of applying the synergistic potential among diverse modalities.
%
% 3) \textit{Incorporating LLM enhances recommendation effectiveness.}
% %
% Models augmented with LLMs (BLIP-2, QWen2-VL, NoteLLM2-NE) consistently outperform those that do not employ LLMs,
% %
% including both ID-based and modality-integrated approaches. 
% %
% This illustrates that the advanced reasoning capabilities and extensive knowledge stored in LLMs can significantly boost recommendation performance.
%
3) \textit{\RAMP~exhibits superiority among all baselines.}
Even our naive version, \ie \modelname\textbf{-NA}, exceeds the best-performed LLM-based baseline, \ie NoteLLM2-NE on Grocery and Sports.
First, compared to classical MSBR without LLMs, \RAMP~achieves outstanding results by fully capitalizing on LLM semantic modeling to generate comprehensive representations. 
Second, \RAMP~also beats these advanced LLM-enhanced approaches, attributed to factors:
i) Unlike BLIP-2 and QWen2-VL which merely model intrinsic item features, \RAMP~decouples transition-aware and knowledge-reflected features with parallel MLLMs, aligning them into a unified semantic space.
ii) Compared with NoteLLM2-NE which utilizes sample-level contrastive loss to extract low-order relations, our model distills transitional patterns into multimodal learning, thus producing more informative representations for downstream tasks.

\subsection{Ablation Studies (RQ2)}

\subsubsection{\textbf{Study of Transfer-MLLM}} 
% 1) 模块loss消融 
We compare three variants:
(a) \textbf{w/o IntraCL-t} removes intra-order contrastive learning on texual features.
(b) \textbf{w/o IntraCL-v} removes contrastive learning on visual features.
(c) \textbf{w/o Gen-T} removes the generation loss in $T$-$MLLM$.
Results in Table \ref{tab:ablation} indicate:
1) Removing any part leads to a decline in performance, and \textbf{w/o Gen-T} performs worst among three variants.
We infer that the $\mathcal{L}_{Gen}$ plays the most critical role in extracting precise multi-order transitions.
2) \modelname~beats \textbf{w/o IntraCL-v} in all metrics, but the gap is not very large, meaning optimizing visual features offers limited overall improvement.
We reach a similar conclusion in \textit{Section 4.4.4}
3) \textbf{w/o IntraCL-t} performs worse than \textbf{w/o IntraCL-v}. Textual contrastive learning enhances the uniformity of similar intra-order patterns and ensures discrimination among different patterns, thus promoting representation learning.

\subsubsection{\textbf{Study of Knowledge-MLLM}}
% 2) 知识loss消融 done 
We test four variants:
(a) \textbf{w/o MIL-h} removes alignment $\mathcal{L}^h_{MIL}$ between collaborative features and transitional patterns.
(b) \textbf{w/o MIL-l} removes alignment $\mathcal{L}^l_{MIL}$ between inherent features and zero-order features.
(c) \textbf{w/o MIU} removes the feature disentanglement $\mathcal{L}_{MIU}$
(d) \textbf{w/o GEN-K} removes the generation loss $\mathcal{L}_{Gen}$ in $K$-$MLLM$.
From Table \ref{tab:ablation}:
1) \textbf{w/o Gen-K} performs worst among all variants, indicating that the generation task dominates the optimization of LLM learning, thus affecting the representation quality most significantly.
2) \modelname~outperforms \textbf{w/o MIU} and \textbf{w/o MIL-h}, which proves necessity of both feature disentanglement and alignment.
3) Although \textbf{w/o MIL-h} performs better than pure DIMO, it shows comparable accuracy with $K$-$MLLM^{(0)}$, because no collaborative signals are incorporated without $\mathcal{L}^h_{MIL}$.
4) \textbf{w/o MIL-l} is inferior to \modelname, proving that overly aligning with collaborations without enhancing knowledge-reflected features also degrades embedding quality.

\subsubsection{\textbf{Effects of Transition Distillation}}
We compare \modelname~with \modelname\textbf{-NA} which removes the transition distillation stage.
% (a) \textbf{$K$-$MLLM^{(0)}$} extracts only item inherent knowledge-reflected features.
% %
% (b) \textbf{$K$-$MLLM^{(1)}$} aligns with transitional patterns based on $T$-$MLLM$.
% %
% (c) \modelname~aligns with the second-order patterns based on $T$-$MLLM^{(2)}$.
%
In Table \ref{tab:compare}, \modelname~significantly outperforms \modelname\textbf{-NA}, showing distillation of transitional patterns into LLM learning depict more informative multimodal representations so as to promote MSBR inferring user interests.
We show ablation results of distilling different-order transition pairs into $K$-$MLLM$ in Appendix C.

% We present the performance of downstream recommendation tasks in Table \ref{tab:stage-backbone}, from which: 
% %
% 1) \modelname~outperforms both $K$-$MLLM^{(0)}$ and $K$-$MLLM^{(1)}$, showing distillation of multi-order patterns into LLM learning depict more informative multimodal representations so as to promote MSBR inferring user interests.
% %
% 2) The accuracy gap between $K$-$MLLM^{(1)}$ and $K$-$MLLM^{(0)}$ is more obvious than that between \modelname~and $K$-$MLLM^{(1)}$, which indicates low-order collaborative signals provide stronger enhancement to features than high-order signals, aligning with common sense that short-term interests dominate pattern trends in SBR while time intervals weaken long-term associations.

\begin{figure}[t]
\vspace{-0.05 in}
\centering
\includegraphics[width=1.0\linewidth]{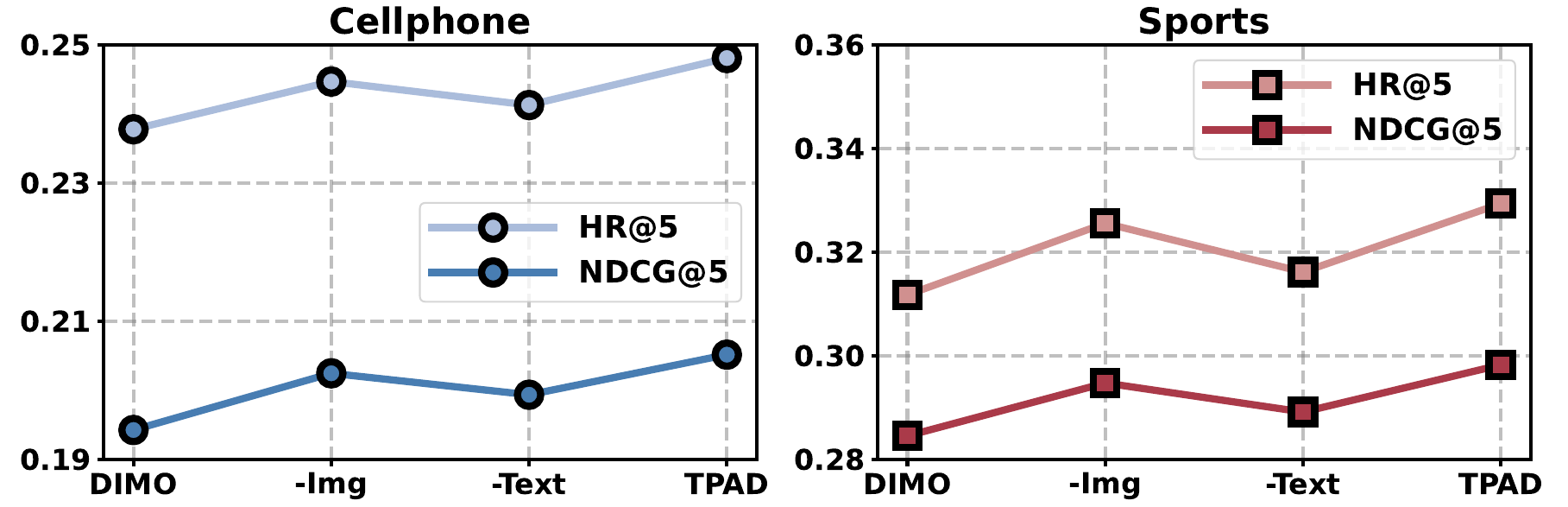}
\vspace{-0.15 in}
\caption{Ablations on different modalities.} 
\label{fig:modality}
\vspace{-0.1 in}
\end{figure}

\subsubsection{\textbf{Effects of Different Modalities}}
% 4）模态消融（画图）
We showcase two variants to investigate the degree of representation promotion in each modality:
(a) \textbf{-Img} denotes that we utilize the textual embeddings from \modelname~in downstream DIMO and maintain visual embeddings derived from pre-trained models.
(b) \textbf{-Text} denotes that we only utilize the visual embeddings from \modelname~in downstream DIMO.
We report the comparison results in Fig \ref{fig:modality}, from which we conclude:
There is a common relationship of \modelname~>\textbf{-Img}>\textbf{-Text}>\textbf{DIMO} on both datasets.
\modelname~exhibits superiority in modeling both modalities, but the contribution of visual and textual modal varies markedly.
\textbf{-Text} performs worse than \textbf{-Img}, which can be attributed from two aspects:
1) For recommender datasets, item textual attributes (\eg title and brand) directly reflect user-diverse intents, yet visual features (\eg shape and style) are relatively subtle thus having weaker associations with user behavior.
2) Although we align visual embeddings into the token space, LLMs are still more adept at processing textual language than handling transformed visual knowledge \cite{zhang2024notellm-2}.
How to further balance learning bias of modalities in large language models remains a future research question.

\begin{figure}[t]
\centering
\includegraphics[width=1.0\linewidth]{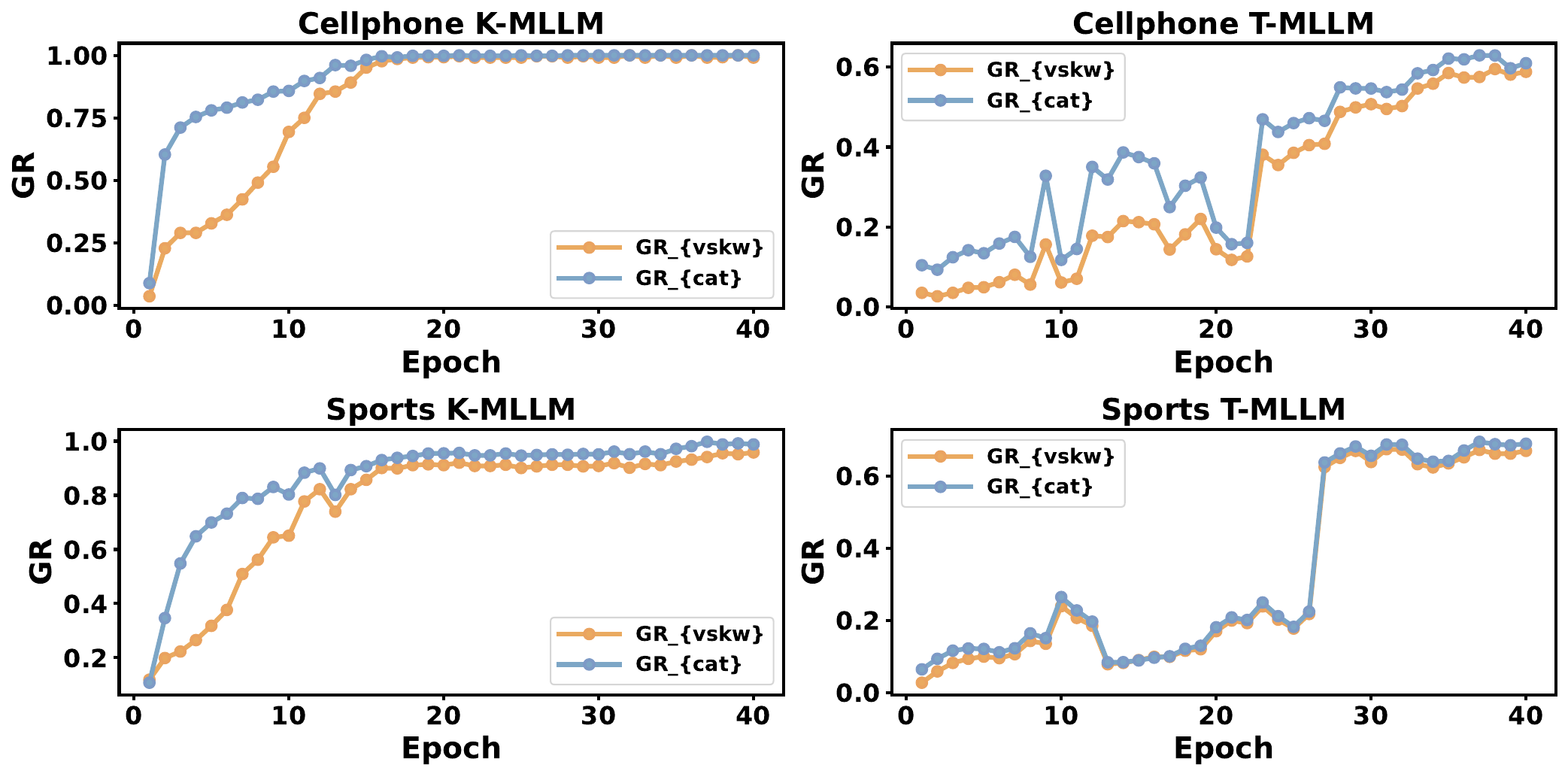}
% \vspace{-0.25 in}
\caption{Performance of generation tasks.} 
\label{acc}
\vspace{-0.1 in}
\end{figure}

\subsection{In-depth Analysis (RQ3)}

\subsubsection{\textbf{Generation Task Performance}}
To assess the training of \modelname, we propose a new metric $GR=p_{mc}/p_{all}$ for measuring the quality of generation results from $T$-$MLLM$ and $K$-$MLLM$, where $p_{mc}$ is the number of generated answers matching the ground truth and $p_{all}$ denotes the prompt amount.
We have $GR_{cat}$ and $GR_{vskw}$ to represent the category and visual keyword prediction accuracy. 
We report the inference accuracy of the final state of two MLLMs in Fig. \ref{acc}: 1) For \textit{K-MLLM}, either $GR_{cat}$ or $GR_{vskw}$ approaches 1.0 with about 15 training epochs, proving the effectiveness of capturing both item textual and visual features.
2) For \textit{T-MLLM}, $GR$ shows a fluctuating upward trend, with several notable breakthrough points, which suggests (i) predicting the next item's features is much harder than summarizing a single item content.
(ii) The process of LLM capturing content collaborations has shifted from simple to complex pattern learning, and breakthrough points mean LLM graspes higher-order global patterns, leading to more generalized predictions across various samples.
% 视觉预测和文本预测的准确率变化是同步的，视觉准确率会低于文本准确率
3) In both MLLMs,  $GR_{cat}$ or $GR_{vskw}$ displays synchronized changes, indicating the balanced learning across both modalities in \modelname.
The slightly lower $GR_{vskw}$ reflects the greater challenge of extracting visual features and the LLM's natural adaptation to text-based tasks.

\begin{table}[t]
\small
\centering
\caption{Inference time (s).}
\label{inference}

\resizebox{1.0\linewidth}{!}{
\begin{tabular}{c|cc|cc}
\toprule
\textbf{Prompts}
& \textbf{Qwen2-VL}
& \textbf{NoteLLM2-NE}
& \modelname\textbf{-NA}
& \modelname \\

\midrule
\textbf{1}  & \textbf{0.352} & \textbf{0.076}  & \textbf{0.069*} & \textbf{0.070}\\

\textbf{100}  & \textbf{34.81} & \textbf{7.134}  & \textbf{6.785*} &  \textbf{6.851}  \\

\textbf{1,000} & \textbf{343.9} & \textbf{68.45} &\textbf{ 65.31*} & \textbf{66.06}\\
\bottomrule

\end{tabular}
%}
}
% \vspace{-0.05 in}
\end{table}

\begin{figure}[t]
% \vspace{-0.05 in}
\centering
\includegraphics[width=1.0\linewidth]{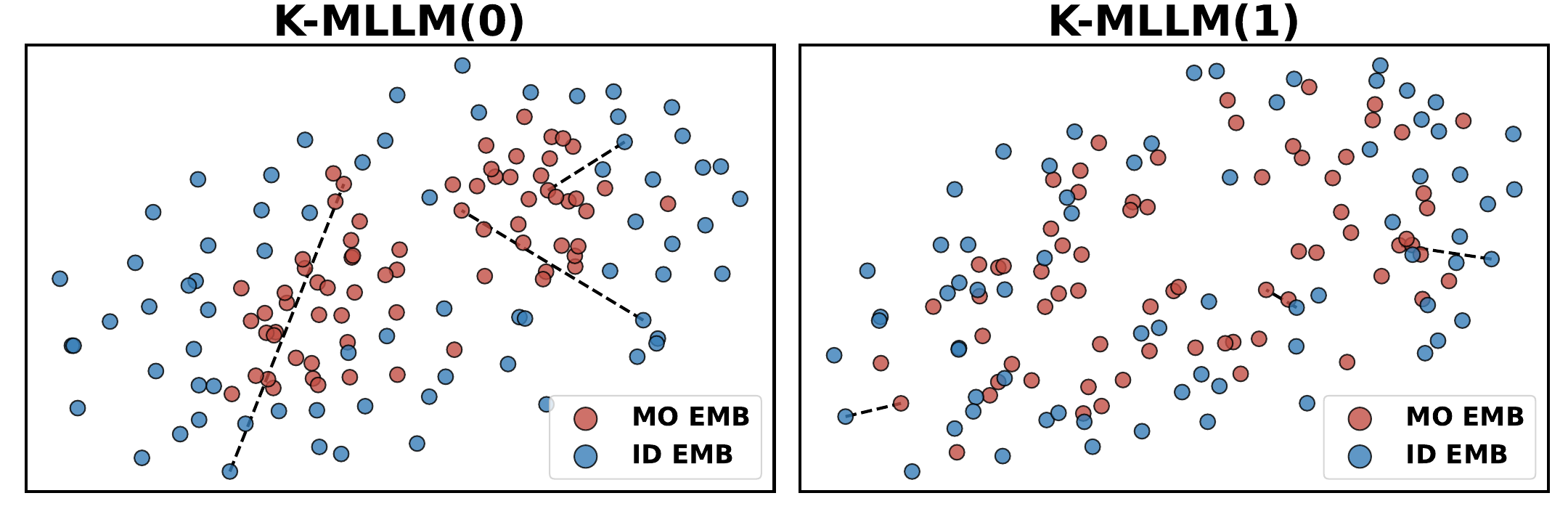}
% \vspace{-0.25 in}
\caption{The t-SNE results of ID and modality feature pairs.} 
\label{tsne}
\vspace{-0.05 in}
\end{figure}

\subsubsection{\textbf{Inference Efficiency}}
We test the inference time for 1, 100, and 1,000 prompts on \modelname~and LLM-based baselines.
Results in Table \ref{inference} show that \modelname\textbf{-NA} and \modelname~have comparable inference speed which is much faster than Qwen2-VL and NoteLLM2-NE.
Compared with NoteLLM2-NE, our model enhances multimodal representations by transitional distillation in the training stage and reduces computational cost of late fusion in inference.
Qwen2-VL show slower inference in our scenarios due to the added complexity of directly processing both raw visual inputs through separate encoders and fusion mechanisms.

% all three MLLMs exhibit relatively fast inference speeds.
%
% \textit{T-MLLMs} are slightly slower than \textit{K-MLLM} because their prompts are longer, including historically clicked items.
%
% $T$-$MLLM^{(2)}$ displays a comparable speed with $T$-$MLLM^{(1)}$, which proves our recurrent prompts facilitate transition inference from low-order to high-order at a stable speed. 

\subsubsection{\textbf{Alignment Interpretability}}
To verify whether TPA module reduces discrepancy between inherent features and transitional patterns, we apply t-SNE to visualize the distributions between ID and modality embeddings.
We pre-train the downstream MSBR with only session data to obtain ID embeddings.
In Fig \ref{tsne}, we visualize the sampled ID embedding distribution with the textual embeddings generated by each state of \textit{K-MLLM}.
Results show:
1) $K$-$MLLM^{(0)}$ extracts only knowledge-reflected features from items, so the distribution bias between ID and modality is significant.
2) $K$-$MLLM^{(1)}$ mitigates distribution discrepancies since it aligns with transition-aware features from $T$-$MLLM$, which proves effectiveness of the TPA module.
%
% 3) \modelname~exhibits highest degree of distribution consistency, reflecting the constraints of higher-order signal alignment on the modality cognition of LLMs.
%
But distribution bias between modality and ID still remains, which offers valuable information in downstream modeling.

\subsubsection{\textbf{Case Study}}
We present case studies on the generation results of \modelname~in Appendix D.

\subsubsection{\textbf{Parameter Analysis}}
We present the parameter analysis and detailed setting in Appendix E.
% We vary $\mu$ and $\gamma_1$ from 0.001 to 0.05, $\gamma_2$ and $\gamma_3$ from 0.1 to 10 in Fig \ref{para}.
% %
% We observe: 1) As we gradually increase $\mu$, the accuracy first boosts and then decreases, indicating the $\mathcal{L}_{\operatorname{IntraCL}}$ enhances intra-order pattern learning, but a huge $\mu$ will disrupt the language task learning of LLM.
% %
% 2) The performance improves with $\gamma_1$ raises and declines when $\gamma_1$ is too large, suggesting the quality of latent feature decoupling is influenced by the weight parameters.
% %
% 3) The bell-shape curves of $\gamma_2$ and $\gamma_3$ show that controlling an appropriate intensity of alignment between transitional patterns realizes distilling transition-aware features into embeddings while preserving multimodal knowledge.

\begin{table}[t]
\centering
\caption{Performance on different MSBR backbones.}
% \vspace{-0.4cm}
\label{tab:stage-backbone}
\resizebox{\linewidth}{!}{
\begin{tabular}{c|cc|cc}
\toprule
\multirow{2.2}{*}{\textbf{Variants}}
& \multicolumn{2}{c|}{\textbf{Grocery}}
& \multicolumn{2}{c}{\textbf{Sports}}\\
\cmidrule(lr){2-3} \cmidrule(lr){4-5} 
&HR@5 &HR@10 &HR@5 &HR@10\\
\midrule
\textbf{MMSBR} &0.5038&	0.5309& 0.3067&	0.3311\\
$\textbf{Qwen2-VL}_{+MMSBR}$  &0.5105	&0.5397	&0.3123	&0.3365	\\
$\textbf{NoteLLM2-NE}_{+MMSBR}$ 	&0.5137	&0.5425	&0.3168	&0.3407	\\
\rowcolor{gray!18} 
$\modelname_{+MMSBR}$ &\textbf{0.5281} &\textbf{0.5527}	&\textbf{0.3279}	&\textbf{0.3512}\\

\bottomrule

\end{tabular}
%}
}
% \vspace{-0.10 in}
\end{table}

\begin{figure}[t]
\centering
% \vspace{-0.10 in}
\includegraphics[width=1.0\linewidth]{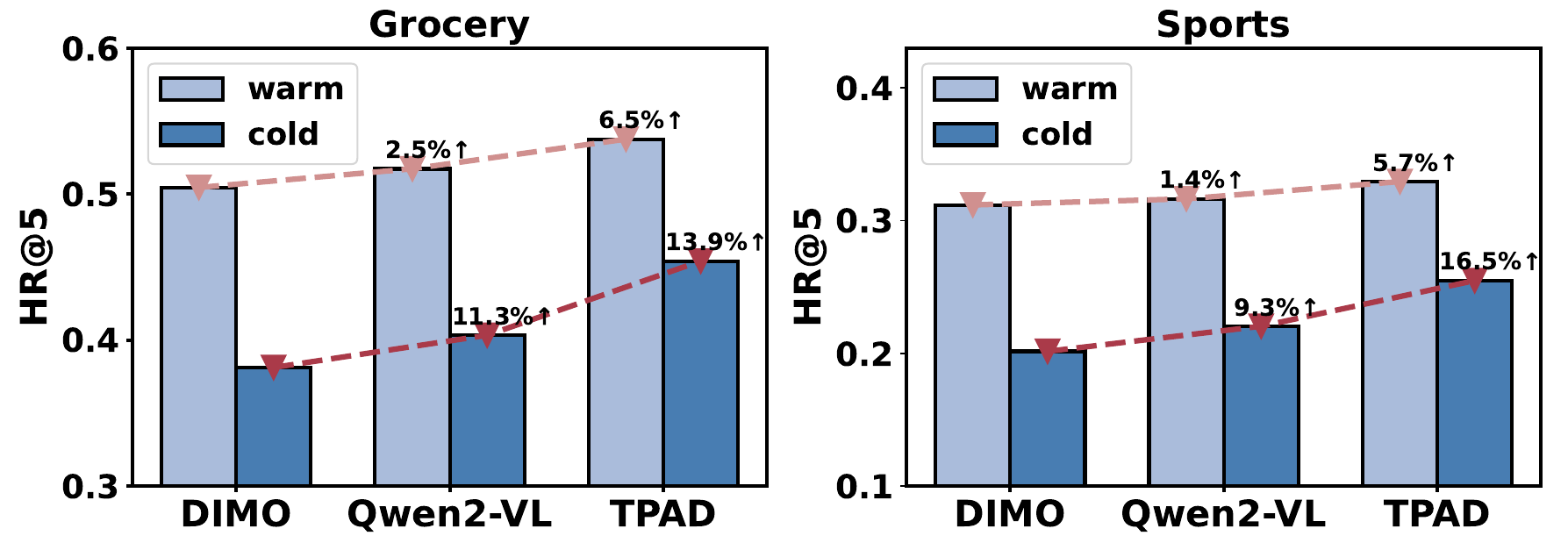}
% \vspace{-0.25 in}
\caption{Performance in warm and cold scenarios.} 
\label{cold_start}
\vspace{-0.1 in}
\end{figure}

\subsection{Handling Various Scenarios}

\subsubsection{\textbf{\textit{Adapt to Different MSBRs}}}
To validate the robustness of our model with different downstream recommenders, we plug \modelname~into another MSBR SOTA, \ie MMSBR \cite{zhang2023beyond}.
We compare performance of LLM-based baselines when adapting to MMSBR, results are shown in Table \ref{tab:stage-backbone}.
We observe that \modelname~substantially boosts the precision of MMSBR, verifying robustness of multimodal embeddings generated by $K$-$MLLM$ in various downstream models.
Our method also outperforms both Qwen2-VL+ and NoteLLM2-NE, showing that distillation on transitional patterns strengthens the LLM comprehension of item context semantics so as to provide more \textbf{adaptive} representations applicable in diverse models.

\subsubsection{\textbf{\textit{Cold-start Setting}}}
% Incorporating multi-modalities can enhance SBR cold-start capability since relations among \textit{cold items} can be inferred from modal features.
Multi-modalities enhance cold-start SBR by inferring relations among \textit{cold items} from modal features.
We refilter datasets to reserve 30\% of cold items in the testing set, which have never appeared in training. 
Fig \ref{cold_start} shows all models face an accuracy drop in the cold scenario, indicating only modal associations are insufficient to portray user intents.
\modelname~has the least accuracy degradation, outperforms baselines by a large margin in cold-start.
Compared to DIMO, \modelname~shows a 6.5\% and 5.7\% improvement on the warm \textit{Grocery} and \textit{Sports} but 13.9\% and 16.5\% in the cold setting.
This phenomenon reveals that \modelname~facilitates distilling transitional patterns into item multimodal representations, promoting LLM's comprehensive cognition on cold items.

\section{Conclusion and Future Work}

We proposed an LLM-enhanced MSBR framework \modelname~that develops a distillation paradigm of integrating transitional patterns into multimodal representation generation.
\modelname~contained two stages, \ie the Knowledge Acquisition Stage and Transition Distillation Stage.
The first stage extracted knowledge-aware features by instructing Knowledge-MLLM with item modality reasoning prompts and multimodal supervised fine-tuning.
The second stage exploited transitional patterns through promoting Transfer-MLLM, and distilled transition-aware features into LLM comprehension.
Extensive experiments demonstrated the effectiveness of \modelname.
This work marks an initial step in leveraging LLM to advance MSBR, while leaving future work as: 
i) Introduce more modalities into LLM while ensuring the efficiency of joint modeling.
ii) Harness feedback from downstream recommenders to guide LLM multimodal generation, establishing a paradigm of preference optimization.

% \begin{acks}
% This work was supported in part by the National Key
% R\&D Program of China (No. 2022YFB4501504).
% \end{acks}

\bibliographystyle{ACM-Reference-Format}
\balance
\bibliography{sample-base}

% \clearpage

% \appendix
% \input{./chapter/appendix.tex}

\end{document}